\documentclass[usenatbib,usegraphicx]{mn2e}
\usepackage{times}
\usepackage{amssymb}

\usepackage[total={17.8cm,24.0cm},centering]{geometry}

\newcommand{\atlas}{{ATLAS$^{\rm 3D}$}}
\newcommand{\Sauron}{\texttt{SAURON}}

\title[The \atlas\ project - XII. $M/L$ recovery of barred galaxies.]{The \atlas\ project - XII. Recovery of the mass-to-light ratio of simulated early-type barred galaxies with axisymmetric dynamical models}

\author[P.-Y. Lablanche et al.]{Pierre-Yves Lablanche$^{1,2}$\thanks{E-mail:plablanc@eso.org},
Michele Cappellari$^3$,
Eric Emsellem$^{2,1}$, \and
Fr\'ed\'eric Bournaud$^4$,
Leo Michel-Dansac$^1$,
Katherine Alatalo$^6$,
Leo Blitz$^6$,\and
Maxime Bois$^7$,
Martin Bureau$^3$,
Roger L. Davies$^3$,
Timothy A. Davis$^3$,\and
P. T. de Zeeuw$^{2,8}$,
Pierre-Alain Duc$^4$,
Sadegh Khochfar$^9$,
Davor Krajnovi\'c$^2$,\and
Harald Kuntschner$^{2}$,
Raffaella Morganti$^{10,11}$,
Richard M. McDermid $^{12}$,\and
Thorsten Naab$^{13}$,
Tom Oosterloo$^{10,11}$,
Marc Sarzi$^{14}$,
Nicholas Scott$^3$,
Paolo Serra$^{10}$,\and
Anne-Marie Weijmans$^{15}$\thanks{Dunlap Fellow}
and Lisa M. Young$^5$\\
$^1$Universit\'e Lyon 1, Observatoire de Lyon, Centre de Recherche Astrophysique de Lyon\\
and Ecole Normale Sup\'erieure de Lyon, 9 avenue Charles Andr\'e, F-69230 Saint-Genis Laval, France\\
$^2$European Southern Observatory, Karl-Schwarzschild-Str. 2, 85748 Garching, Germany\\
$^3$Sub-department of Astrophysics, Department of Physics, University of Oxford, Denys Wilkinson Building, Keble Road, Oxford OX1 3RH\\
$^4$Laboratoire AIM Paris-Saclay, CEA/IRFU/SAp  CNRS  Universit\'e Paris Diderot, 91191 Gif-sur-Yvette Cedex, France\\
$^5$Physics Department, New Mexico Institute of Mining and Technology, Socorro, NM 87801, USA\\
$^6$Department of Astronomy, Campbell Hall, University of California, Berkeley, CA 94720, USA\\
$^7$Observatoire de Paris, LERMA and CNRS, 61 Av. de l'Observatoire, F-75014 Paris, France\\
$^8$Sterrewacht Leiden, Leiden University, Postbus 9513, 2300 RA Leiden, the Netherlands\\
$^9$Max Planck Institut f\"ur extraterrestrische Physik, PO Box 1312, D-85478 Garching, Germany\\
$^{10}$Netherlands Institute for Radio Astronomy (ASTRON), Postbus 2, 7990 AA Dwingeloo, The Netherlands\\
$^{11}$Kapteyn Astronomical Institute, University of Groningen, Postbus 800, 9700 AV Groningen, The Netherlands\\
$^{12}$Gemini Observatory, Northern Operations Centre, 670 N. A`ohoku Place, Hilo, HI 96720, USA\\
$^{13}$Max-Planck-Institut f\"ur Astrophysik, Karl-Schwarzschild-Str. 1, 85741 Garching, Germany\\
$^{14}$Centre for Astrophysics Research, University of Hertfordshire, Hatfield, Herts AL1 9AB, UK\\
$^{15}$Dunlap Institute for Astronomy \& Astrophysics, University of Toronto, 50 St. George Street, Toronto, ON M5S 3H4, Canada }

\begin{document}

\date{Accepted 2012 May 18. Received 2012 May 16; in original form 2011 October 5}

\pagerange{\pageref{firstpage}--\pageref{lastpage}} \pubyear{2012}

\maketitle

\label{firstpage}

\clearpage
\begin{abstract}
We investigate the accuracy in the recovery of the stellar dynamics of barred galaxies when using axisymmetric dynamical models.
We do this by trying to recover the mass-to-light ratio ($M/L$) and the anisotropy of realistic galaxy simulations using the Jeans Anisotropic Multi-Gaussian Expansion (JAM) method.
However, given that the biases we find are mostly due to an application of an axisymmetric modeling algorithm to a non-axisymmetric system and in particular to inaccuracies in the de-projected mass model, our results are relevant for general axisymmetric modelling methods.
We run N-body collisionless simulations to build a library with various luminosity distribution, constructed to mimic real individual galaxies, with realistic anisotropy.
The final result of our evolved library of simulations contains both barred and unbarred galaxies.
The JAM method assumes an axisymmetric mass distribution, and we adopt a spatially constant $M/L$ and anisotropy $\beta_z=1-\sigma_z^2/\sigma_R^2$ distributions.
The models are fitted to two-dimensional maps of the second velocity moments $V_{\rm rms}=\sqrt{V^2+\sigma^2}$ of the simulations for various viewing angles (position angle of the bar and inclination of the galaxy).
We find that the inclination is generally well recovered by the JAM models, for both barred and unbarred simulations.
For unbarred simulations the $M/L$ is also accurately recovered, with negligible median bias and with a maximum one of just $\Delta (M/L)<1.5$\% when the galaxy is not too close to face on.
At very low inclinations ($i\la30^\circ$) the $M/L$ can be significantly overestimated (9\% in our tests, but errors can be larger for very face-on views).
This is in agreement with previous studies.
For barred simulations the $M/L$ is on average (when PA$=45^\circ$) essentially unbiased, but we measure an over/under estimation of up to $\Delta (M/L)=15$\% in our tests.
The sign of the $M/L$ bias depends on the position angle of the bar as expected: overestimation occurs when the bar is closer to end-on, due to the increased stellar motion along the line-of-sight, and underestimation otherwise.
For unbarred simulations, the JAM models are able to recover the mean value of the anisotropy with bias $\Delta\beta_z\la0.1$, within the region constrained by the kinematics.
However when a bar is present, or for nearly face-on models, the recovered anisotropy varies wildly, with biases up to $\Delta\beta_z\approx0.3$.
\end{abstract}


\section{Introduction}

The determination of the masses (or equivalently mass-to-light ratios) of gas-poor galaxies has been an important issue since the discovery that galaxies are stellar systems like the Milky Way, with mass being a strong driver of many of their properties.
Dynamical modelling methods of increased sophistication have been developed over the past decades, all based on the assumption that galaxies can be described as stationary systems.
The first attempt at measuring dynamical masses of galaxies were based on the spherical virial equations \citep{Pov58,Spit69}.
These methods have the disadvantage that, for accurate results, they need to assume self-similarity in the galaxy light and mass distribution.
More accurate methods allow for axisymmetry and take the galaxy light distribution into account.
The first detailed axisymmetric models of real galaxies were based on the \cite{Jeans22} equations and assumed a distribution function that depends on two (out of three) integrals of motion \citep[e.g][]{Bin90,vdm90,Em94b}, but special classes of three-integral models were also used.
Axisymmetric methods were developed to allow for a general orbital distribution, based on \cite{Schwa79} numerical orbital superposition method \citep[e.g.][]{Cret99,vdm98,Geb03,Thom04,Cap06}.
Currently the most general available models assume galaxies can be approximated by a stationary triaxial shape \citep[e.g.][]{del07,vdb08}.

The above modelling techniques were developed under the assumption that gas-poor galaxies can be well described by stationary axisymmetric or triaxial spheroidal systems.
However a key initial result of the \atlas\ survey \citep[][hereafter Paper~I]{Cap11} is the fact that nearby gas-poor galaxies are actually dominated (86 per cent of them) by fast rotators \citep[][hereafter Paper~II and Paper~III]{Kraj11,Em2011}, often with significant disk components and resembling spiral galaxies with the dust removed \citep[][Paper VII]{CapP7}, 30\% of which at least are barred.
The presence of these bars is a difficult problem for all modelling methods and therefore motivates the present study.

Bars are density waves which results in a tumbling potential: this figure rotation is often ignored in the popular dynamical
modelling methods described above. Dynamical models of barred galaxies have been constructed in the past \citep[e.g.][]{Pfe84,Haf00,Zhao96}.
However, the existence of intrinsic degeneracies in the dynamical modelling of bars make the determination of mass quite uncertain even for such models. In fact even the full amount of information one can obtain today for external galaxies, namely the full line-of-sight velocity distribution (LOSVD) at every position on the sky, is not sufficient to uniquely constrain the two free parameters ($M/L$ and inclination) of a simple self-consistent axisymmetric model \citep{Val04, Kraj05, Cap06, vdb09}.
A barred model requires at least two extra parameters (the Position Angle (PA) and pattern speed of the bar) and dramatically increases the complexity of the orbital structure and the associated degeneracy of the problem, instead of improving the accuracy of the mass estimate: a broad range of parameters space may well fit the data equally well.
Moreover, assuming a galaxy is barred also increases the degeneracy in the mass deprojection problem \citep[e.g.][]{Ger96}, which is already mathematically non unique in the simple axisymmetric case \citep{Ryb87}.
The application of sophisticated barred models to large samples would be computationally challenging, but feasible exploiting the trivial parallelism of the problem.
However, this brute-force approach does not remove the intrinsic degeneracies so it is not expected to increase the accuracy of the mass determinations, and for this reason does not seem justified.

An alternative approach consists of using some a priori information on the galaxy structure and make empirically-motivated restrictive assumption on the models.
This is the approach we are using in the systematic determination of the masses of the 260 early-type galaxies of the \atlas\ survey \citep{Cap2012}.
We are applying the Multi-Gaussian Expansion (MGE) technique \citep{Em94} to accurately describe the photometry of all galaxies in the survey (Scott et al. in prep) and use the Jeans Anisotropic MGE (JAM) modelling method \citep{Cap08} to measure masses.
The JAM method is based on a simple and very efficient solution of the Jeans equations which allows for orbital anisotropy $\beta_z=1-(\sigma_z/\sigma_R)^2$.
This approach provides good descriptions of the integral-field kinematics of the fast rotator early-type galaxies \citep{Cap08,Sco09,Cap2012}, which constitute the large majority of the \atlas\ sample (see Paper~II and Paper~III).
A key motivation for our use of the JAM method is that it was shown, using 25 real galaxies \citep{Cap06,Cap08}, to agree well within the model uncertainties in the mass determination obtained with the more general axisymmetric Schwarzschild approach.

The use of an axisymmetric dynamical modelling method to measure the mass of barred galaxies raises the obvious question of what errors in the mass determination are introduced by the approach.
Answering this question is the goal of this paper.
Our work is in its spirit an extension to barred disk galaxies of the work by \cite{Thom07} , which explored the biases introduced by the use of axisymmetric models, when extracting masses of triaxial and prolate simulated spheroidal galaxy remnants.

The paper is structured as follows: in Section~\ref{sec:Model} we briefly describe the Multi-Gaussian Expansion parametrization from \cite{Em94}, the semi-isotropic Jeans equations and the method used to create the N-body galaxy models. In Section~\ref{sec:JAMmodel}, we give an overview of the input models used for the JAM modeling in our investigation, and in Section~\ref{sec:results} we compare the original and recovered values
of the corresponding dynamical parameters. Section~\ref{sec:Sum} summarizes all results.

\section[]{Modeling and N-body Simulations of early-type barred galaxies}\label{sec:Model}

\subsection{Mass modeling}\label{sec:MGE}

\subsubsection{Multi-Gaussian Expansion method} \label{sec:MGEf}

\begin{figure*}
\includegraphics[width=2\columnwidth]{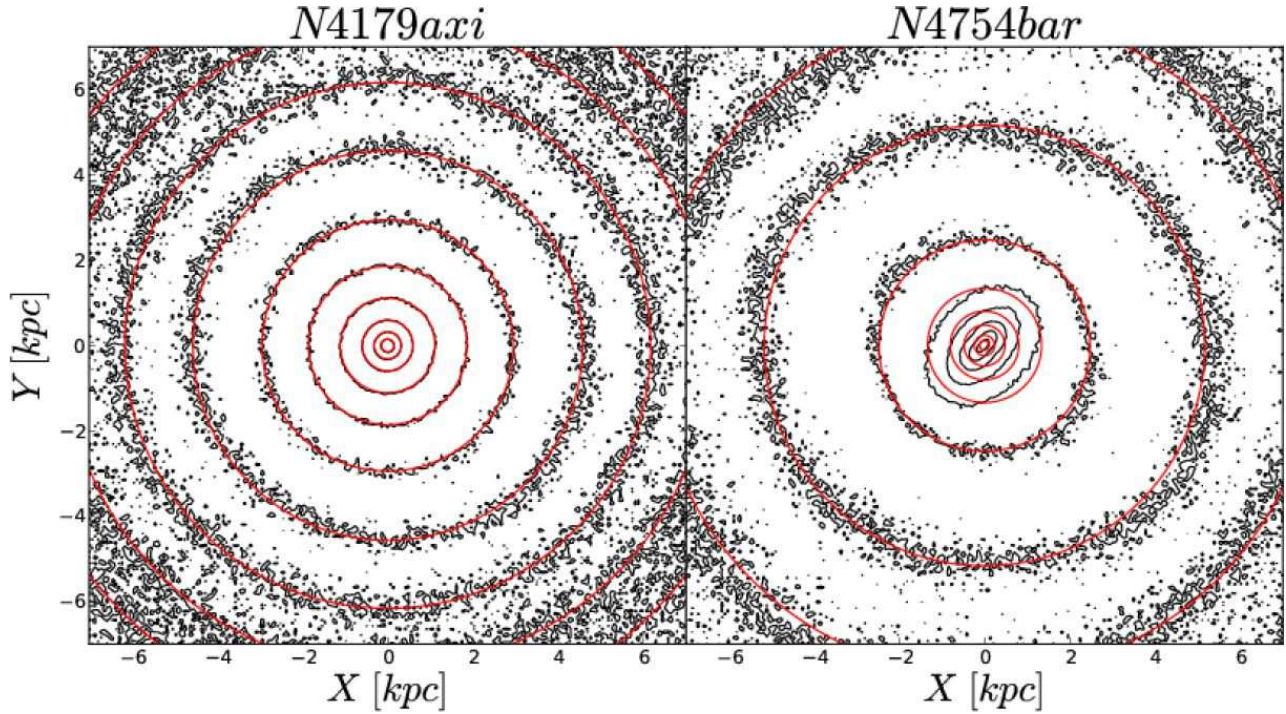}
\caption{Examples of MGE fits to two model galaxies with a projection angle of $i=25^{\circ}$. The left panel shows the MGE fit to simulated galaxy N4179axi, an axisymmetric object. On the right we show the fit to N4754bar with PA$_{\rm bar}=60^{\circ}$, where the bar was excluded from the fit. \label{fig:MGEs}}
\end{figure*}

We use in our study the Multi-Gaussian Expansion method described in \citet{Em94} and \citet{Cap02}.
The technique basically consists of decomposing the luminosity into a number of concentric two-dimensional (2D) Gaussians. By fitting the detailed surface brightness distribution, the MGE formalism provides a description of the intrinsic luminosity density, which converts to the mass distribution via the assumed constant $M/L$.
From galaxy images this formalism allows us to generate realistic initial conditions for our N-body simulations using the method explained in Sec~\ref{sec:Nbody}.
The MGE parametrization is also the first and crucial step of the JAM modeling.
Thus, a rigorous and robust approach is needed when producing the MGE model of a galaxy, as the predicted kinematics may significantly depend on the obtained mass distribution (see Sec~\ref{sec:JAM}).
The method and software\footnote{available from http://purl.org/cappellari/idl} we adopt in our study to produce MGE parametrization is fully described in \cite{Cap02}.

Once the best fit has been found, we have a description of the galaxy surface brightness distribution given as a sum of two-dimensional Gaussians which we can attempt to deproject.
The deprojection of a galaxy surface brightness distribution is formally non-unique for all but edge-on cases, and the degeneracy can become severe at low inclinations \citep{Ryb87}.
The MGE method provides just \textit{one} solution for the deprojection, in terms of
a sum of three-dimensional Gaussians. This method has been intensively used and
usually provides luminosity distributions consistent with observed photometry of
existing galaxies, but the MGE method obviously does not remove the existing intrinsic degeneracy.

The deprojection of an MGE model can be done analytically once the viewing angles are known \citep[see][]{Monnet92}.
When the system is assumed to be axisymmetric, only one viewing angle, the inclination $i$
($i=90^{\circ}$ for an edge-on system), is sufficient to retrieve the full three-dimensional luminosity distribution $\nu$
(if the galaxy is not face-on).

In a coordinate system $(x',y',z')$ centered on the galaxy nucleus with \textit{z'} pointing toward us and $(x',y')$ being the plane of the sky, the MGE surface brightness can be written as:
\begin{equation} \label{eq:surf}
   \Sigma(x',y') = \sum_{k=1}^{N}\frac{L_k}{2\pi\sigma_k'^2q_k'} \exp{\left[-\frac{1}{2\sigma_k'^2}\left({x'}_r^2+\frac{{y'}_r^2}{q_k'^2}\right)\right]}
\end{equation}
where N is the number of adopted gaussian components, each having an
integrated luminosity $L_k$, an observed axial ratio $0 \leqslant q_k' \leqslant 1$,
a dispersion $\sigma_k'$ along the major axis, and a position angle (PA) $\psi_k$ measured counter-clockwise from $y'$ to the major-axis of the Gaussian, with
$(x'_r, y'_r)$ being the associated rotated coordinate system.
Then, the deprojected MGE luminosity distribution in cylindrical coordinates can be expressed as:

\begin{equation} \label{eq:nu}
   \nu(R,z) = \sum_{k=1}^N\frac{L_k}{(\sqrt{2\pi}\sigma_k)^3q_k} \exp{\left[-\frac{1}{2\sigma_k^2}\left(R^2 + \frac{z^2}{q_k^2}\right)\right]}
\end{equation}
where the $k$-th gaussian has the total luminosity $L_k$, intrinsic axial ratio $q_k$ and dispersion $\sigma_k$ (In the present study $\sigma_k = \sigma_k'$) The intrinsic axial ratio $q_k$ can then be written as

\begin{equation}\label{eq:q}
   q_k^2 = \frac{{q'}_k^2 - \cos^2{i}}{\sin^2{i}}, \,\,\, {\rm for }\,\, i \ne 0
\end{equation}
where $i$ is the galaxy inclination.

\subsubsection{MGE modeling of barred and unbarred S0 galaxies}

The fitting of the MGE to the photometry follows the procedure applied by \cite{Sco09} (see their Fig.~2) to deal with the presence of bars.
This same approach is being applied to the MGE fits of the \atlas\ sample (Scott et al. in preparation).
The procedure allows to find the best fit to the photometry, maximizing the minimum $q'_k$ and minimizing the maximum $q'_k$, while still being consistent with the projected galaxy image within the errors.
For near face-on cases, a small deviation in the observed axis ratios implies an important change in the flatness (or roundness)
of the mass distribution.
As the previous procedure frequently ends with very similar lower and upper limits for $q'_k$, we force a common axis ratio for all Gaussians for such cases to keep an acceptable global shape for our MGE model.

When the bar clearly affects the projected photometry of our models (i.e., when
it is easily detected), the method adopted for the MGE parametrization consists in forcing the lower and upper limit of the gaussian axis ratios.
In this context, a bar can be considered as a perturbation of a disk structure.
Bars, if fully fitted by MGE components, appear as Gaussians elongated along the apparent long axis of the bar.
The presence of a bar tends to significantly affect the $q'_k$ values of a few
Gaussians depending on its position angle, strength and length (the position angle of the
bar PA$_{\rm bar}$ being measured counter-clockwise from the projected major-axis of the galaxy).
The resulting $q'_k$ could thus make the system look flatter (or rounder) than the corresponding axisymmetric case if the bar is seen end-on (resp. side-on).
Previous tests made in \cite{Sco09} showed that the best fitting MGE parametrization of a barred galaxy is usually not the one which allows the best fit to the observed kinematics (using JAM models).
The kinematic fit is significantly improved when the Gaussians have constrained axial ratios such that the systems is forced to an axisymmetric ''bar-less'' MGE parametrization.
As bars often only affect the photometry within a restricted radial range, we use the outer disk of each galaxy to constrain the imposed value of
the Gaussian flattening.
Figure~\ref{fig:MGEs} gives two examples of the resulting MGE fits for an axisymmetric simulation and a barred simulation, both including large-scale disks.

\begin{table*}
 \begin{tabular}{ c c c c c c c c c}
  \hline
  \hline
  Model & Galaxy/Model & Distance (Mpc) & $N_{*}$ & $\beta_{zi}$ & $\beta_{zf}$ & $\sigma_{\phi}/\sigma_{R}$ & time(Gyr) & bar \\
  \hline
  \hline
  N4179axi & NGC4179 & 16.5 & 4e6 & $\beta(\varepsilon)$ & 0.106 & 1.8 & 1.5 & no \\
  N4570axi & NGC4570 & 17.1 & 4e6 & 0.0 & 0.145 & 1.0 & 1.5 & no \\
  \\
  N4442bar & NGC4442 & 15.3 & 4e6 & $\beta(\varepsilon)$ & 0.344 & 1.8 & 1.5 & yes \\
  N4754bar & NGC4754 & 16.1 & 4e6 & $\beta(\varepsilon)$ & 0.343 & 1.8 & 1.5 & yes \\
  \hline
  \hline
 \end{tabular}
 \centering
 \caption{Table providing a list of simulations with their labels and their specifications. Distances are set according to Paper~I. $N_{*}$ is the number of particles. $\beta_{zi}$ gives the anisotropy of the initial conditions as described in Section~\ref{subsec:DynStru}. $\beta_{zf}$ is the global anisotropy computed for the final state of our simulations. $\sigma_{\phi}/\sigma_{R}$ gives the second relation for the geometry of the velocity dispersion ellipsoid for the initial conditions. The time given in the eighth column correponds to the simulated time of evolution. The last column indicates whether a bar appeared or not in our simulations. \label{tab:model_simu}}
\end{table*}

\subsection{Jeans Anisotropic MGE Modeling} \label{sec:JAM}

The JAM method is a powerful approach to model the stellar kinematics of early-type galaxies,
providing a good description of the first two stellar velocity
moments ($V$, $V_{\rm rms}$) of a stellar system. This technique can be used to probe the dynamical structure of ETGs and does in principle allow the recovery of the inclination and the dynamical mass-to-light $M/L$ ratio.
The JAM technique allows for a different M/L and anisotropy for each individual MGE Gaussian component.
However, the measurement of a global mass for real galaxies does not seem to require this extra generality, at least within 1$R_e$, where good quality integral-field data are available \citep[e.g.][and Paper~I]{Em04}. For this reason the models we use make the following simple assumptions \citep[a full description of JAM is provided in][]{Cap08}:
\begin{enumerate}
 \item An axisymmetric distribution of the mass.
 \item A constant mass-to-light $M/L$ ratio.
 \item A constant anisotropy described by the classic anisotropy parameter $\beta_z = 1 - (\sigma_z/\sigma_R)^2$  with $\beta_z \gtrsim 0$.
\end{enumerate}

When the mass distribution is represented via an MGE parametrization (see
Sect.~\ref{sec:MGE} above), the Jeans equations can be easily integrated
along the line-of-sight as shown by \cite{Em94} in the semi-isotropic case ($\sigma_z = \sigma_R \neq \sigma_{\phi}$) and by \cite{Cap08} in the anisotropic generalization ($\sigma_z \neq \sigma_R \neq \sigma_{\phi}$).
Here we use the anisotropic formulas (equations 28 and 38 of \cite{Cap08}) to derive the projected first and
second velocity moments ($V$ and $V_{\rm rms}$) given a set of input parameters (MGE mass model,
mass-to-light ratio $M/L$, anisotropy $\beta_z$), and thus find the best fitting
values within a sampled predefined solution space (e.g., $\beta_z \ge 0$)

Such a JAM method has been systemically applied to \Sauron\ integral-field stellar kinematics
of all 260 early-type galaxies of the \atlas\ sample. In the present study, we
rely on mock observations computed from N-body simulations of galaxies, and we chose
to build JAM models from artificial maps to mimic the procedure used in the course
of the \atlas\ survey. We also exclude the central few arcseconds during the fitting process,
which e.g., avoids biases due to the effect of the seeing.
We also re-bin all maps before fitting by using the Voronoi tessellation as described in \cite{Vor}:
this allows a guaranteed minimum signal-to-noise ratio in each bin, and reduces the
scatter in the outer parts of the kinematic maps.


\subsection{N-body simulations of regular-rotator galaxies}\label{sec:Nbody}

\begin{table}
 \begin{tabular}{ c c c c c c c }
  \hline
  \hline
  Model & Galaxy/Model & $q$ & Dist & $N_{*}$ & $\beta_{zi}$ & $\frac{\sigma_{\phi}}{\sigma_{R}}$ \\
  \hline
  \hline
  Hern01 & Hernquist & 1.0 & 10.0 & 2e6 & 0.0 & 1.0 \\
  Hern02 & Hernquist & 1.0 & 10.0 & 2e6 & 0.2 & 1.0 \\
  Hern03* & Hernquist(flat) & 0.5 & 10.0 & 2e6 & 0.0 & 1.0 \\
  Hern04* & Hernquist(flat) & 0.5 & 10.0 & 2e6 & 0.2 & 1.0 \\
  N4754ini & NGC4754 & 0.4-0.2 & 16.1 & 4e6 & 0.2 & 1.0 \\
  \hline
  \hline
  \end{tabular}
 \centering
 \caption{Table providing a list of the tests simulations (static models) with their labels and their specifications. Distance (in Mpc) of $N4754ini$ was set according to Paper~I, while for Hernquist models it was set arbitrary. The flattening of the models is given through the axis ratio $q$ of the MGE models (for N4754ini axis ratio cover a range from $q=0.4$ for inner gaussians to $q=0.2$ for outer ones). $N_{*}$ is the number of particles. $\beta_{zi}$ gives the anisotropy of the initial conditions as described in Section~\ref{subsec:DynStru}. $\sigma_{\phi}/\sigma_{R}$ gives the second relation for the geometry of the velocity dispersion ellipsoid for the initial conditions. \label{tab:model_test}}
\end{table}

As the main motivation of our study is to find the influence of a bar on the recovery of basic dynamical
parameters with the JAM method, we chose to use an N-body approach to generate simulations of barred early-type galaxies:
knowing the exact input dynamics for these simulations, we can then compare the key
parameters with those determined via the JAM modeling.
We also made static realisations of a few (Hernquist and one typical axisymmetric lenticular) mass models, and thus only used the initial realisation
of the N-body distribution. These models are detailed in Table~\ref{tab:model_test}, while details for evolved simulations are summarized in Table~\ref{tab:model_simu}.
The method to build the initial conditions for our simulations (to be evolved,
or not) is detailed below.

\subsubsection{Particle positions}

Starting from the MGE parametrization of a mass distribution (after taking into account the mass-to-light ratio $M/L$), the initial positions of the particles can be computed easily.
Each Gaussian represents a fraction of the total mass, so that given a total number of particles per component we can determine $N_k$, the number of particles of that $k$-th Gaussian.
All components are truncated at a chosen radius.
To set up the position of each particle, we use a standard realisation method with a random generator via the cumulative function of a (truncated) Gaussian function, scaling each spatial dimension with the corresponding spatial dispersion.

\subsubsection{Dynamical structure}\label{subsec:DynStru}
Given the particle position, we compute the velocity dispersion components $\sigma_R$, $\sigma_{\phi}$ and $\sigma_z$ solving Jeans Equations \citep{Jeans22}, within the MGE formalism of \cite{Em94}.
For this work we use the anisotropic generalization of the method (equations 19--21 and 34 of \cite{Cap08}), which allows one to set arbitrary ratios $\sigma_z/\sigma_R$ and $\sigma_{\phi}/\sigma_R$ for the axes of the velocity ellipsoid, which is assumed to be cylindrically oriented.
Values for these ratios can be set individually for each gaussian component, but in this study all Gaussians share the same geometry of its velocity dispersion ellipsoid.
For some of the simulations initial conditions, we used the following (so-called $\beta-\epsilon$ , with $\epsilon$ the intrinsic galaxy ellipticity) relation to fix $\sigma_R / \sigma_z$:
\begin{equation} \label{eq:betaeps}
 \beta_{z} = 1 - \left(\frac{\sigma_z}{\sigma_R}\right)^2 = 0.6 \times \epsilon
\end{equation}
This a purely empirical relation which seems to describe the general trend in the anisotropy of real fast rotator galaxies \citep{Cap07}.

\subsubsection{Simulation code}
The numerical simulations are performed with a particle-mesh N-body code \citep{Bournaud07}.
The density is computed with a Cloud-in-Cell interpolation, and an FFT-based Poisson solver is used to compute the gravitational potential, with a spatial resolution and softening of 48 pc.
Particle motions are integrated with a leap-frog algorithm and a time-step of 0.1 Myr.
The number of particles and the time evolution of each model are given in Table~\ref{tab:model_simu}.

\section{Inputs for JAM models}\label{sec:JAMmodel}


\begin{figure*}
 \includegraphics[width=2\columnwidth]{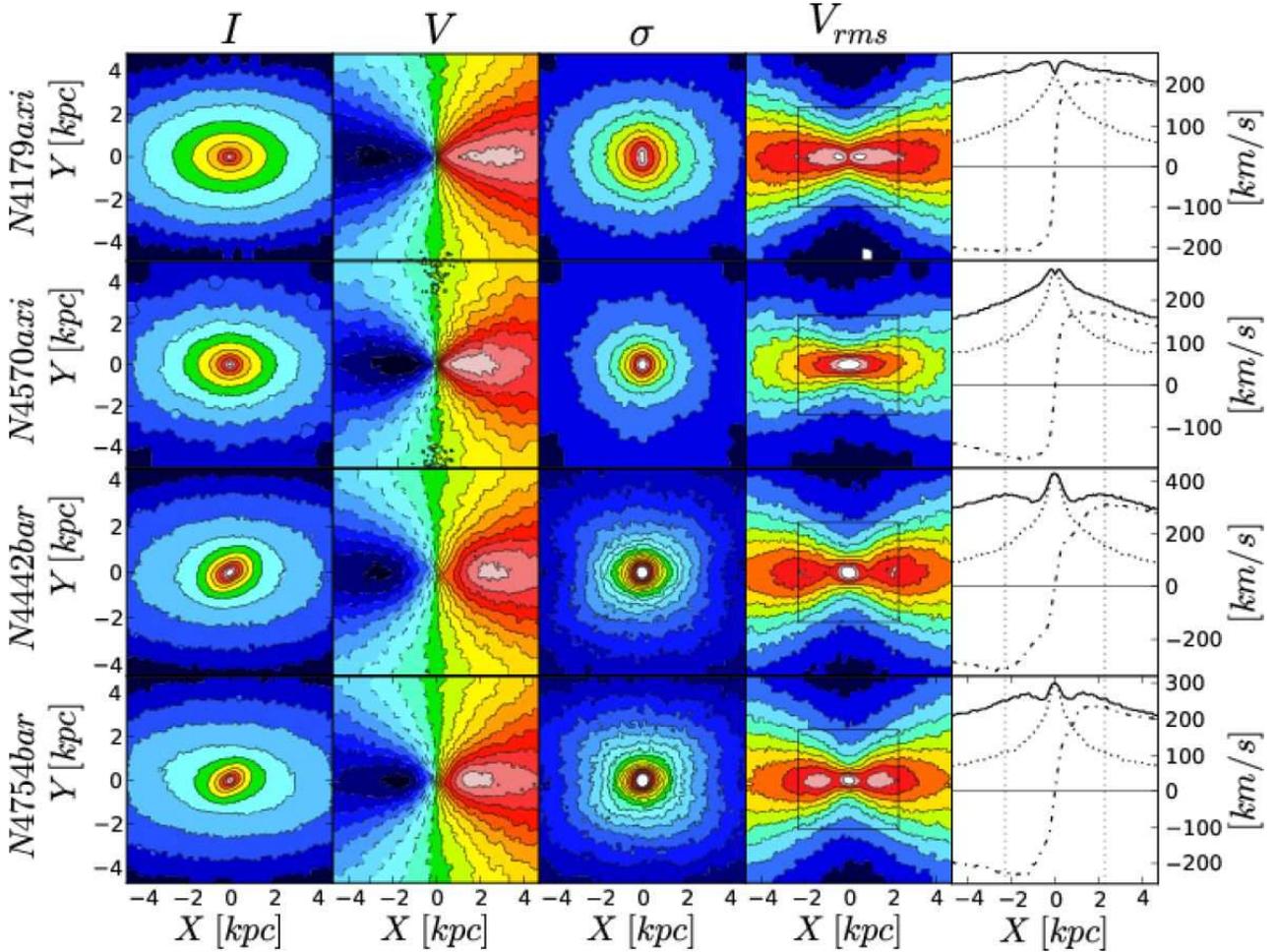}
\caption{Binned map of the projections of the final state of our simulations with an inclination $i=60^{\circ}$ and PA$_{\rm bar}=60^{\circ}$ for barred simulations. From left to right : total projected luminosity $I$ ; mean velocity along the line-of-sight $V_{\rm LOS}$ ; velocity dispersion along line-of-sight $\sigma_{\rm LOS}$ ; projected second velocity moment $V_{\rm rms} = \sqrt{V_{\rm LOS}^2 + \sigma_{\rm LOS}^2}$ ; profiles of $V_{\rm LOS}$ (dot line), $\sigma_{\rm LOS}$ (dash-dot line) and $V_{\rm rms}$ (solid line) along the galaxy major axis. From top to bottom : $N4179axi$ ; $N4570axi$ ; $N4442bar$ ; $N4754bar$.The square box in the $V_{\rm rms}$ map and the vertical dot lines in the profiles represent the typical field of view used in our study. \label{fig:Panel}}
\end{figure*}

\subsection{Unbarred simulations}\label{sec:axi}

We first built four simple simulations based on the analytic \cite{Hern90} mass distribution
\begin{equation}
   \rho(r) = \frac{M}{2\pi} \frac{a}{r} \frac{1}{(r+a)^3}
\end{equation}
where $M$ is the total mass and $a$ is a scale length.
These simulations allowed us to quickly check the purely numerical accuracy of the JAM modeling method and of our starting-conditions generating software for spherical isotropic ($Hern01$), spherical anisotropic ($Hern02$), flat isotropic ($Hern03$) and flat anisotropic ($Hern04$) particle realisations.
Flat Hernquist models are oblate systems derived from the Hernquist profile by forcing an axis ratio of 0.5 for the mass distribution.
For all these simulations the inclination was recovered within an error of less than $2^{\circ}$ (besides the
case of $Hern01$, for which the inclination is meaningless), the global anisotropy was accurately recovered within an error of $\pm 0.025$ and the error on $M/L$ never exceeded $1.5\%$

We then considered a more realistic numerical test, using the MGE parametrization of the SDSS $r$-band image of the real galaxy NGC4754
and a constant anisotropy and we label this model $N4754ini$. When computing the particle velocities,
we forced $\beta_z = 0.2$ for all Gaussians. In contrast to the above-mentioned Hernquist models,
$N4754ini$ represents a complex multi-component object in terms of its mass distribution
and kinematics, and is therefore expected to be more challenging for the JAM modeling method.

We also built two simulations, respectively based on the MGE parametrizations of NGC4570 and NGC4179, which were
evolved via N-body simulations during 1.5Gyr (the face-on and edge-on projections of the final state are
illustrated in \ref{fig:Simu_final}). These two simulations can be considered as fully relaxed, and contrarily to
$N4754ini$ and the Hernquist models, the resulting $\beta_z$ is measured not to be constant with radius.
One important difference between $N4570axi$ and $N4179axi$ is that, while the initial conditions for $N4570axi$
were fixed as isotropic ($\beta_z = 0$), the ones for $N4179axi$ were those of a
dynamically cold disk as described in Table~\ref{tab:model_simu}. No bar formed
either for $N4179axi$ or $N4570axi$.

Projected velocity and velocity dispersion maps of our two axisymmetric simulations with variable
$\beta_z$ are presented in Figure~\ref{fig:Panel}.
The second velocity moment is dominated by the dispersion in the central region, and by the velocity in the outer parts.
The profile of $V_{\rm rms}$ also shows a central depletion.

\begin{figure*}
 \includegraphics[width=\textwidth]{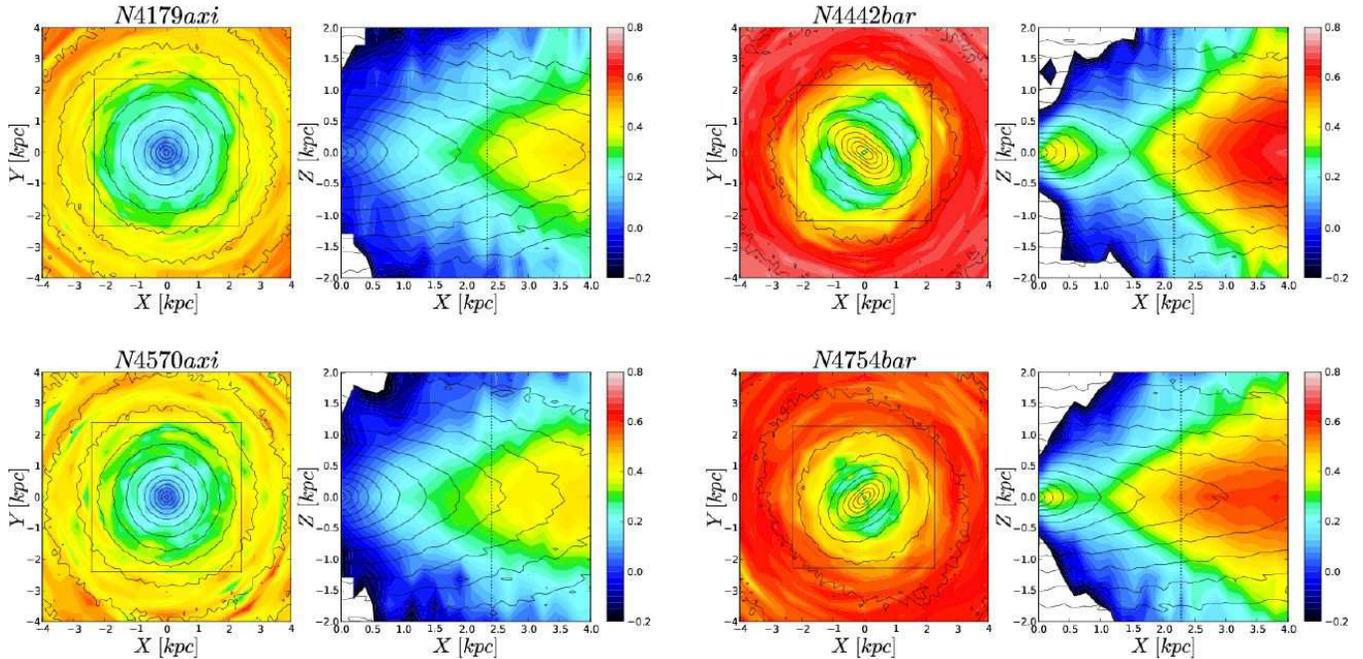}
 \caption{Maps of the local values of the anisotropy $\beta_z$ in the equatorial plane and in the meridional plane of $N4179axi$ (upper left), $N4570axi$ (bottom left), $N4442bar$ (upper right) and $N4754bar$ (bottom right). Colour gradients for the anisotropy go from $\beta_z = -0.2$ (i.e. $\sigma_z > \sigma_R$) to $\beta_z = 0.8$ (i.e. $\sigma_z < \sigma_R$). The field of view used in our study is represented by the solid square for the equatorial plane and by the vertical dot line for the meridional plane. \label{fig:BetaMap}}.
\end{figure*}

Figure~\ref{fig:BetaMap} present the local anisotropy $\beta_z$ as measured
in the meridional plane and the equatorial plane, computed on a cylindrical grid with linearly spaced cells in $R$, in $z$ and in angle $\phi$ where ($R$,$\phi$,$z$) are the standard cylindrical coordinates, including
a central cylindrical cell with a radius of $R_c = 0.01$kpc.
This highlights the fact the anisotropy is not constant in our axisymmetric simulations and that the central parts are more isotropic than the outer regions.
In Figure~\ref{fig:BR}, we show that the radial anisotropy profile (which is simply the azimuthal average of the equatorial computation) generally increases outwards for these axisymmetric
simulations.


\subsection{Barred simulations}\label{sec:bar}
We also developed bar simulations, $N4442bar$, and $N4754bar$, based on the initial MGE parametrizations of NGC4442 and NGC4754, respectively: these are
the final state of two N-body simulations after 1.5~Gyr of evolution (face-on and edge-on projections of the final state can be found in \ref{fig:Simu_final}).
To generate a bar, we force a cold dynamical structure in the initial particle realisations by setting in the initial conditions $\sigma_{\phi} = \sigma_R / 1.8$.
The radial velocity dispersion $\sigma_R$ was set using the $\beta(\varepsilon)$ function described in Eq.~\ref{eq:betaeps}.
With these conditions, a bar appears in each of these simulations after only a few
rotation period, namely between about 25~Myr and 50~Myr of simulated evolution.
As mentioned, we let the galaxy evolve for 1.5~Gyr, to make sure that the bar is well settled.

$N4442bar$ presents the biggest bar (in size) of our simulations: we estimate a semi-major axis of 3.0\,kpc.
The size of the bar is determined using the radial flattening ($q_{isophote} = 1 -\epsilon_{isophote}$) and the position angle ($PA_{\rm isophote}$) of isophotes as done in \cite{WD}.
Basically we define the end of the bar as the radius where isophotes are nearly round ($q_{isophote} > 0.9$) combined with an important change in the position angle.
Using the same method we determine the semi-major axis of the bar for $N4754bar$ to be 2.2~kpc.
Outside of the bar regions, our two simulations are characterized by a rotation pattern consistent with an axisymmetric disk-like system.

The velocity fields of our barred simulations are shown in the two lower panels of Figure~\ref{fig:Panel}.
As for the axisymmetric cases, the $V_{\rm rms}$ maps are dominated by velocity dispersion in the central parts and by the mean velocity in the outer parts.
But in contrast to the $N4179axi$ and $N4754ini$ simulations, the barred simulations all show a peak in the center of the $V_{\rm rms}$ maps.
This apparent difference between the barred and unbarred cases has important consequences, since this will condition the fit of the projected second velocity moment
via a JAM model.

\begin{figure}
 \includegraphics[width=\columnwidth]{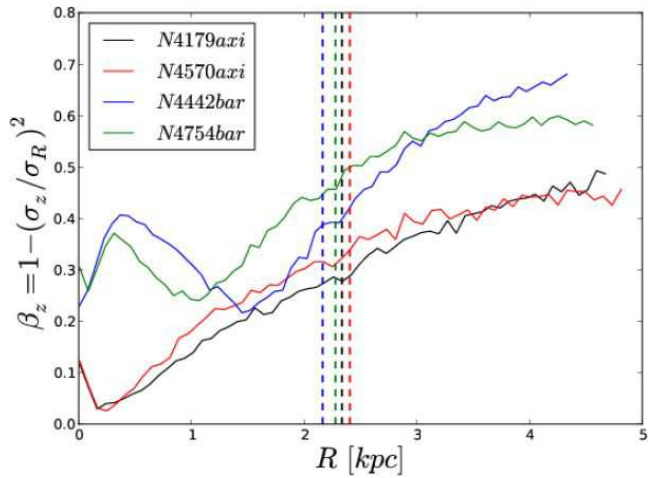}
 \caption{Azimuthally averaged profile of the anisotropy computed in an equatorial plane ($-0.5kpc < z < 0.5kpc$) for axisymmetric simulations ($N4179axi$ and $N4570axi$) and barred simulations ($N4442bar$ and $N4754bar$). The vertical dash-dot lines correspond to the field of view used for the JAM models and therefore the region in which we derived the global anisotropy. \label{fig:BR}}
\end{figure}

Figure~\ref{fig:BetaMap} and Figure~\ref{fig:BR} show that the local anisotropy starts from a central value of 0.3, reaching a minimum of nearly 0.2 close to the end of the bar, then increasing in the outer parts of the model up to 0.7.
The derivation of $\beta_z$ in the equatorial plane confirms the presence of a drop in $\beta_z$ in the outer parts of the bar.


\subsection{Mock observations}\label{sec:mock}

As input for the JAM modeling, we simulated observations by projecting our simulations first with four different inclinations: $i=25^{\circ}$ ; $i=45^{\circ}$ ; $i=60^{\circ}$ and $i=87^{\circ}$,
($i=0^{\circ}$ corresponding to the face-on projection, and $i=90^{\circ}$ to the edge-on projection).
The choice of a near edge-on projection instead of an exact edge-on projection was motivated by the goal of checking the accuracy of the inclination recovery.
Indeed the edge-on projection does not allow an overestimation of $i$ and then limits the range of possible uncertainty.
When a bar is present, we also used four different position angles for the bar,
and this for each value of the inclination: PA$_{\rm bar}=18^{\circ}$ ; PA$_{\rm bar}=45^{\circ}$ ; PA$_{\rm bar}=60^{\circ}$ and PA$_{\rm bar}=87^{\circ}$.
The position angle of the bar is measured counter-clockwise from the galaxy projected major-axis to the bar major-axis, so that PA$_{\rm bar}=87^{\circ}$
is close to having the bar end-on and PA$_{\rm bar}=18^{\circ}$ close to side-on.
We also simulated the SAURON pixels size of $0\farcs8$ assuming a distance for
each model as given in Table~\ref{tab:model_simu} typical of \atlas\ objects.
Note that the $M/L$ provided in Table~\ref{tab:sum} could have been chosen as unity, considering that we will only probe here relative $M/L$.
However, we favoured realistic values as to deal with sensible velocity measurements.

For real galaxies the photometry is available over a much larger field of view than the kinematics.
For each projection, the MGE parametrization was achieved using a rather wide field of view of $401 \times 401$ pixels corresponding to $320.8$~arcsec, and including the full model.
Then, the JAM modeling was fit on a $73 \times 73$ pixels map ($58.4$~arcsec) of the second velocity moment with the simulated galaxy centered on its nucleus, as
this roughly corresponds to the setup for a single SAURON exposure.

\begin{figure}
 \includegraphics[width=\columnwidth]{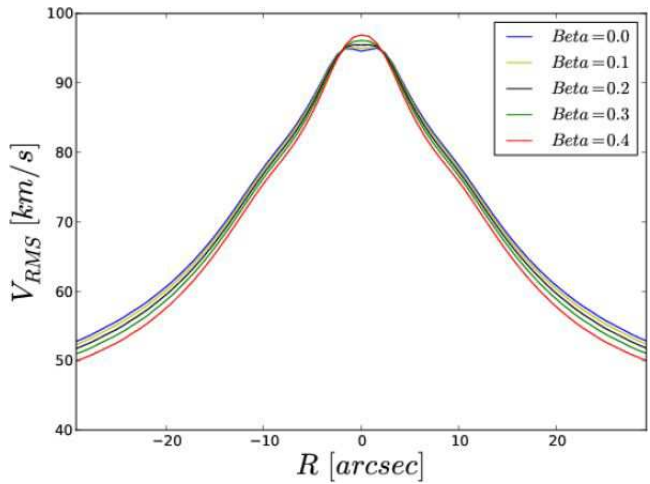}
 \caption{Effect of the anisotropy parameter $\beta_z$ on the observed $V_{\rm rms}$ profile along major axis at an inclination of $i=18^{\circ}$. The difference between all profiles is too small to allow a good recovery of $\beta_z$ at such low inclination. \label{fig:betadeg}}
\end{figure}

\section{Recovery of parameters}\label{sec:results}

The recovery of $i$, $\beta_z$ and $M/L$ with JAM is done on $V_{\rm rms}$ maps only.
The prediction of $V$ and $\sigma$ require an extra assumption on the constancy of the tangential anisotropy of the JAM models.
This assumption may not be well verified in the simulations, especially in barred ones.
However, the accuracy in the determination of the above parameters only depends on the ability of JAM to reproduce the $V_{\rm rms}$, and not the $V$ and $\sigma$ fields separately.
Details of the method to calculate $V$ and $\sigma$ can be found in \cite{Cap08}.


\subsection{Recovery of $\beta_z$ and inclination} \label{sec:beta}

As shown in Figure~\ref{fig:BR} the anisotropy varies significantly when going from the inner to the outer parts of our simulations.

The JAM modelling method allows for a different anisotropy $\beta_z$ for every individual MGE Gaussian.
Here we limit ourselves to the simple case where $\beta_z$ is constant for the whole model.
Tests using 25 real galaxies \citep{Cap06,Cap08} have shown that even with constant anisotropy the recovered $M/L$ agrees with the one derived with Schwarzschild models, which allow for a general anisotropy distribution.
The extra generality of JAM is therefore not required in this case.
Thus, for comparison, we compute a global anisotropy for our simulations by doing a luminosity-weighted average of the local anisotropy for all of our simulations.
In the following we discuss the issue of the inclination-anisotropy degeneracy intrinsic to galaxy dynamics and also the important influence of the MGE parametrization on the recovery of the global anisotropy.
The global anisotropy recovered with the JAM modeling method is noted $\beta_{z}^{\rm JAM}$ and the one computed directly from the simulation $\beta_{z}^{\rm SIM}$.

\subsubsection{Anisotropy-inclination degeneracy}

As shown in \cite{Kraj05} and with more galaxies in \cite{Cap06} there is an intrinsic degeneracy in the dynamical problem between the recovery of the inclination and the anisotropy: given the observed photometry, the observed kinematics can be reproduced in detail for a wide range of inclinations, by varying the orbital make up of the models.
This degeneracy persists even in the restricted case in which the anisotropy is assumed to be constant for the whole galaxy \citep{Cap08}.
The degeneracy can only be broken by making empirically-motivated assumptions on the anisotropy.
This was the approach adopted by \cite{Cap08}, who showed using a small sample of galaxies that, if $\beta_z$ is assumed to be positive, as determined using general models on large samples of galaxies \citep{Cap07,Thom09}, the correct inclination can be recovered from the observed integral-field kinematics.

Here we test the inclination recovery via JAM using N-body simulations of both unbarred (i.e. $N4179axi$ and $N4570axi$) and in particular barred galaxies (i.e. $N4442bar$ and $N4754bar$), for which the inclination is known.
We confirm the fact that, despite the inclination-anisotropy degeneracy the inclination can be recovered in axisymmetric simulations, and we additionally find that the inclination can be recovered even in barred simulations.
The influence of $i$ on the second velocity moment map is important at low inclination.
A slight change in the inclination implies a significant change in $V_{\rm rms}$ (which is more directly related to the change in the axis ratio),
and then allows a good recovery of the inclination value $i$.
At higher inclination, the change in ellipticity is milder and $V_{\rm rms}$ is less influenced by small variations.
In all cases, the uncertainty never exceeds a few degrees, even when the fit can be considered difficult
due to the presence of non-axisymmetric features such as a bar.
We also confirm that at low inclinations ($i\lesssim 40^{\circ}$) there is a degeneracy in anisotropy in the $V_{\rm rms}$ fit, essentially preventing any information on the anisotropy from being extracted from the data in that case.
As shown in Figure~\ref{fig:betadeg}, changes induced by the anisotropy parameter on the $V_{\rm rms}$ profiles are not significant for $i = 18^{\circ}$. This is due to the fact that at low inclination, maps of the second velocity moment are dominated by $\sigma_z$. Therefore, to avoid a too strong degeneracy, $i=25^{\circ}$ is the lowest inclination we will consider in this study. For lower values, the degeneracy prevents $\beta_z$ from being usefully recovered.

To summarize, the higher the inclination, the larger the effect of the global anisotropy on the projected $V_{\rm rms}$.
For the non-barred simulations, following eight different projections (2 galaxies projected at 4 inclinations), we find average errors of 2 degrees and maximum error of 4 degrees at an edge-on view, where the deprojected axial ratio changes most weakly with inclination.
For the barred simulations (32 determinations: 2 galaxies projected at 4 inclinations and 4 PA$_{\rm bar}$) the maximum error in the recovered inclination is 6 degrees.
The maximum error happens only for PA$_{\rm bar} \sim90^{\circ}$, when the bar is seen side-on and it is incorrectly deprojected as a thin disk (see Fig~\ref{fig:MGE_deproj_bis}).
The error on $\beta_{z}^{\rm JAM}$ is important at low inclination and decreases from face-on to edge-on projections.

\begin{figure}
 \includegraphics[width=\columnwidth]{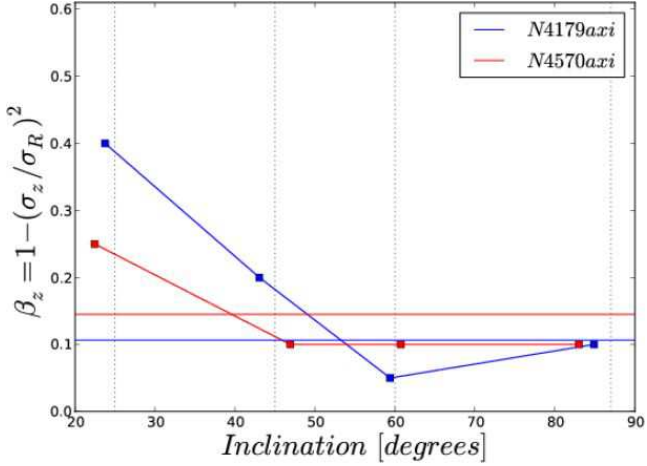}
\caption{Recovered values of the global anisotropy for non-barred simulations as a function of the projection inclination for $N4179axi$ and $N4570axi$. The horizontal lines represent the global anisotropy computed from the particle realisation $\beta_{z}^{\rm SIM}$ (blue for $N4179axi$ and red for $N4570axi$). Vertical lines give the inclination of projections. \label{fig:AxiBeta}}
\end{figure}

\subsubsection{Influence of the mass deprojection}\label{sec:Vsig}

While the inclination-anisotropy degeneracy is intrinsic to the galaxy dynamics and determines the uncertainties for the recovered global anisotropy and the inclination, $\beta_{z}^{\rm JAM}$ is also influenced by the mass deprojection.
For regular axisymmetric simulations such as $N4179axi$ and $N4570axi$, the MGE parametrization is relatively robust, and $\beta_{z}^{\rm JAM}$ is generally very close to the known intrinsic value $\beta_{z}^{\rm SIM}$ (see Figure~\ref{fig:AxiBeta}).
The deviations near face-on view are due to the degeneracy in the mass deprojection at that low inclination.

\begin{figure*}
 \includegraphics[width=2.0\columnwidth]{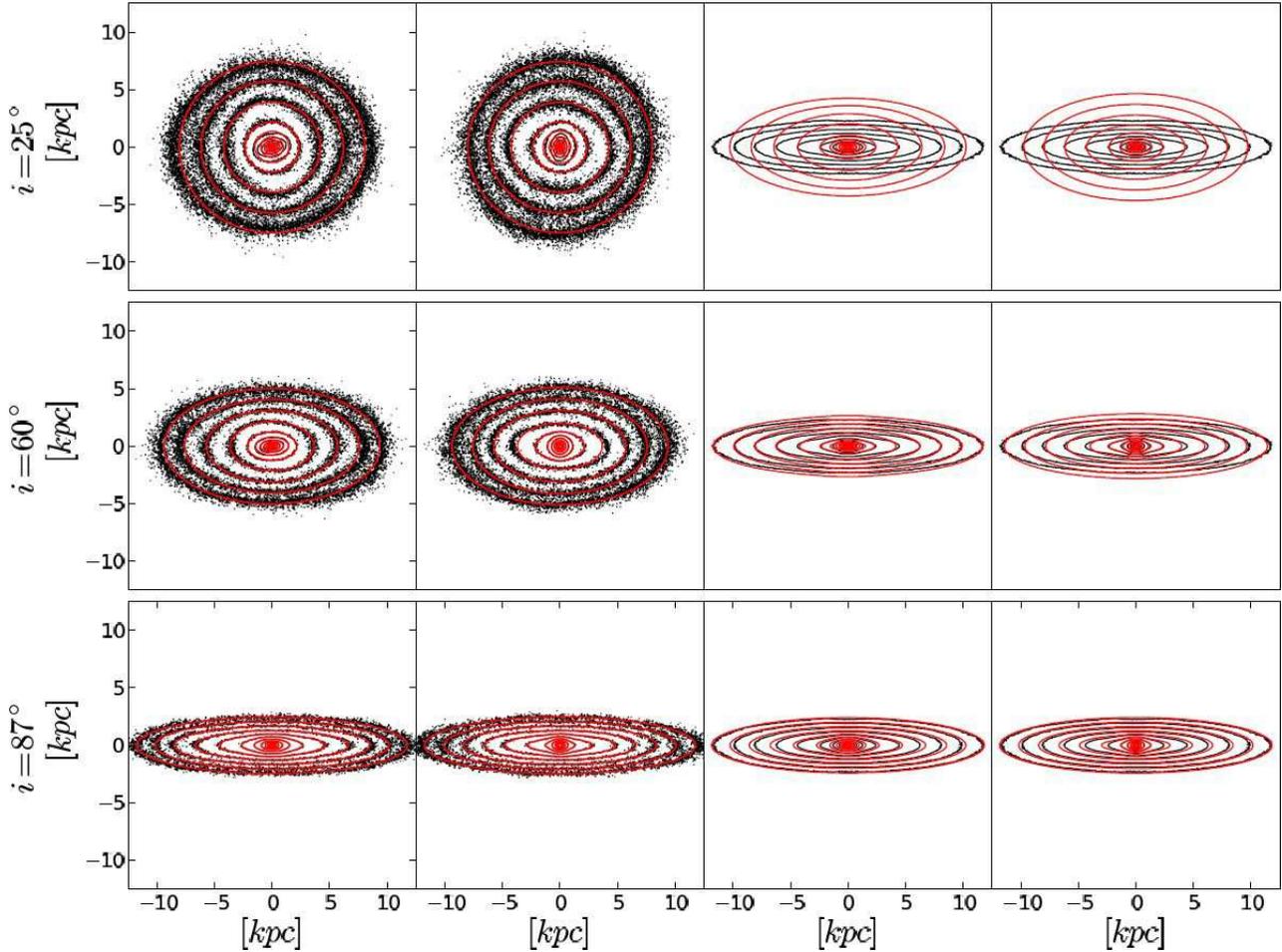}
 \caption{Effect of the inclination on the deprojection of MGE models for $i=25^{\circ}$, $i=60^{\circ}$ and $i=87^{\circ}$. In the first and second columns the MGE fitting (red lines) of the projected mass distribution (black contours) is shown for $PA_{bar}=18^{\circ}$ and $PA_{bar}=87^{\circ}$. The edge-on deprojections of the two latter MGE fitting are represented in the third and fourth columns respectively, superposed to the azimuthally averaged projected density. \label{fig:MGE_deproj}}.
\end{figure*}

\begin{figure*}
 \includegraphics[width=2.0\columnwidth]{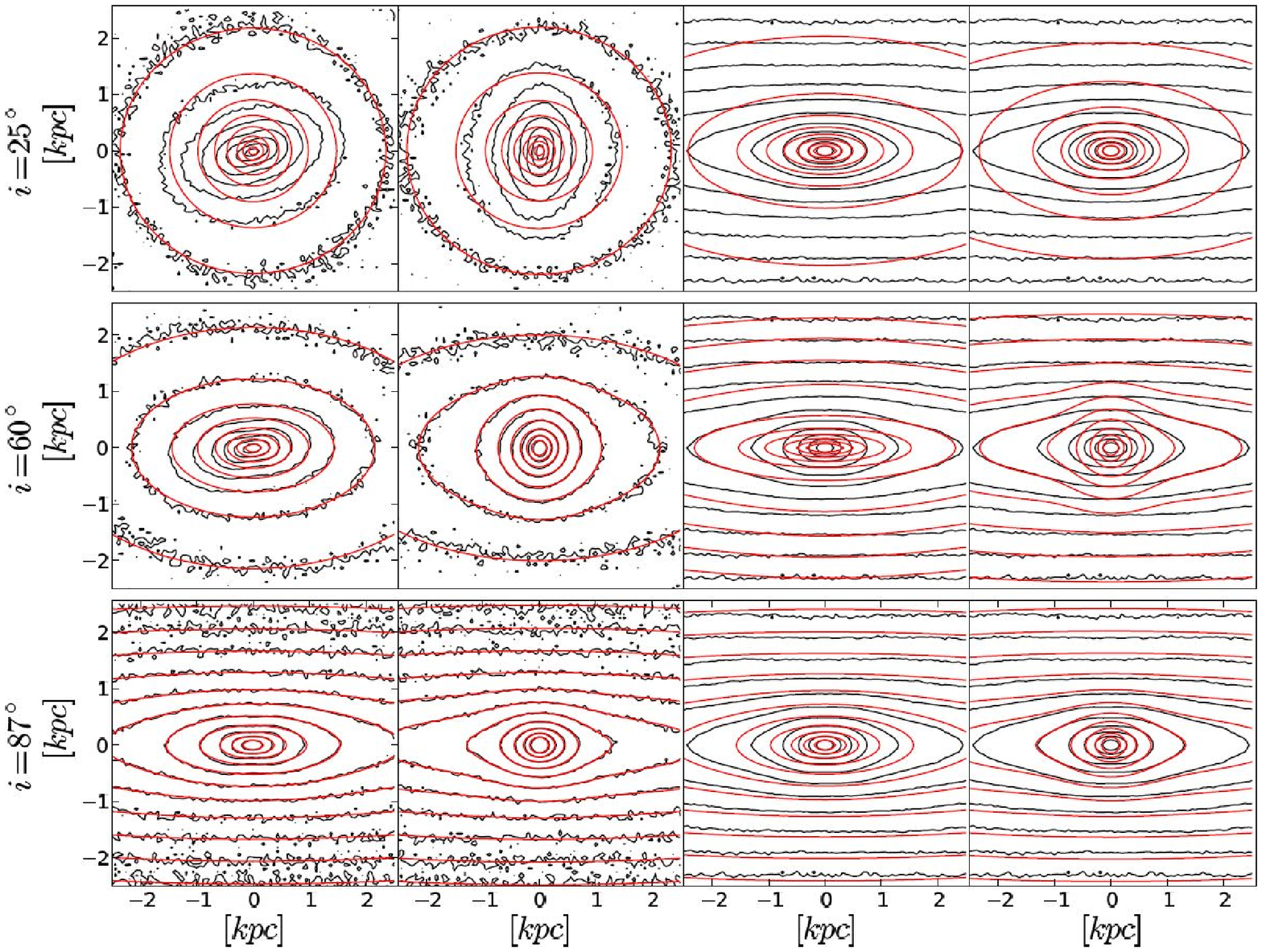}
 \caption{Same as Figure~\ref{fig:MGE_deproj} with a field of view of $5 kpc \times 5 kpc$. \label{fig:MGE_deproj_bis}}.
\end{figure*}

For the barred simulations $N4442bar$ and $N4754bar$, the presence of the bar implies that axisymmetric models cannot reproduce the true stellar density distribution.
This means that our MGE parametrization can only provide an approximation.
Fig~\ref{fig:MGE_deproj} and Fig~\ref{fig:MGE_deproj_bis} illustrate the impact of the degeneracy in
the mass deprojection of barred  galaxies. The deprojected surface brightness MGE model of a barred galaxy
is shown for three inclination of the galaxy ($i=25^{\circ}$, $i=60^{\circ}$ and $i=87^{\circ}$) and
two positions of the bar ($PA_{bar}=18^{\circ}$ and $PA_{bar}=87^{\circ}$).
Close to edge-on ($i = 87^{\circ}$, bottom panels), the reconstructed deprojected models do a reasonable job at fitting the true
edge-on surface bightness contours. When the bar is close to end-on, the impact of assuming an axisymmetric
model becomes more visible. At the other extreme end, near face-on models ($i = 25^{\circ}$, top panels) have deprojected
contours which significantly depart from the true edge-on ones. This is mostly due to the fact that a small
change in the fitted axis ratio of the Gaussians has a large impact on the intrinsic axis ratio  after
deprojection : this is further illustrated and emphasised in Appendix~\ref{fig:MGE_degen}.
For intermediate viewing angles ($i=60^{\circ}$, middle panels), the deprojected photometry fits reasonably
well the true edge-on contours, while again, the discrepancy is emphasised in the region of the bar when it is
initially viewed edge-on. For a real, observed near face-on galaxy, it is hard to know
how close the MGE fitting process would get from the intrinsic axis ratio of the outer disk, as
it would depend on e.g., the regularity of the disk (for example, its lopsidedness),
the signal to noise and contamination from the sky.
but Fig~\ref{fig:MGE_deproj} and Fig~\ref{fig:MGE_deproj_bis}
demonstrate that a significant degeneracy exists when deprojecting targets at low inclination (close to face-on).
These tests show that, as expected, (i) the presence of a bar produces a deprojected density flatter (rounder) that the azimuthally averaged one when the bar is close to side-on (end-on), (ii) near a face-on inclination the deprojected density can be significantly in error.
This obviously impacts the associated dynamical modelling, as detailed in the following Sections.

\begin{figure*}
 \includegraphics[width=\textwidth]{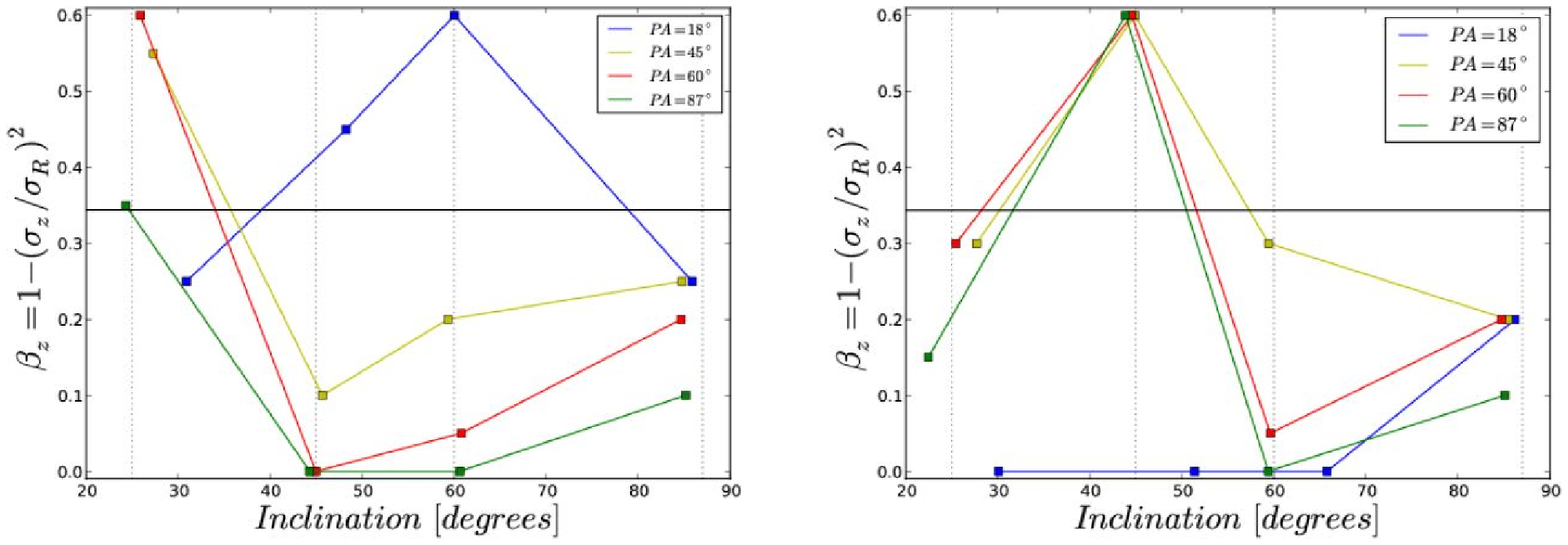}
\caption{Recovered value of the global anisotropy for barred simulations as a function of the projection inclination for $N4442bar$ (left) and $N4754bar$ (right) for different PA$_{\rm bar}$. The horizontal black solid line represent the value of $\beta_{z}^{\rm SIM}$ and the vertical lines shows the inclinations of projection. \label{fig:BarBeta}}
\end{figure*}

The results presented in Figure~\ref{fig:BarBeta} are obtained by applying the same method used on observations.
We can see that the global anisotropy is never well recovered.
This is explained by the fact that the presence of the bar will tend to flatten or round the Gaussians in the MGE parametrization if it is seen respectively side-on (PA$_{\rm bar}=18^{\circ}$) or end-on (PA$_{\rm bar}=87^{\circ}$).
It is important to note that this bias, being due mainly to the mass deprojection (see Fig~\ref{fig:MGE_deproj} and Fig~\ref{fig:MGE_deproj_bis}), is not specific to the adopted dynamical modelling method, but is expected to affect more general method like Schwarzschild's orbit-superposition method.
Moreover the anisotropy is formally not a well defined quantity in a barred galaxy, as it is expected to vary with the azimuthal location on the galaxy disks.

\begin{figure}
 \includegraphics[width=\columnwidth]{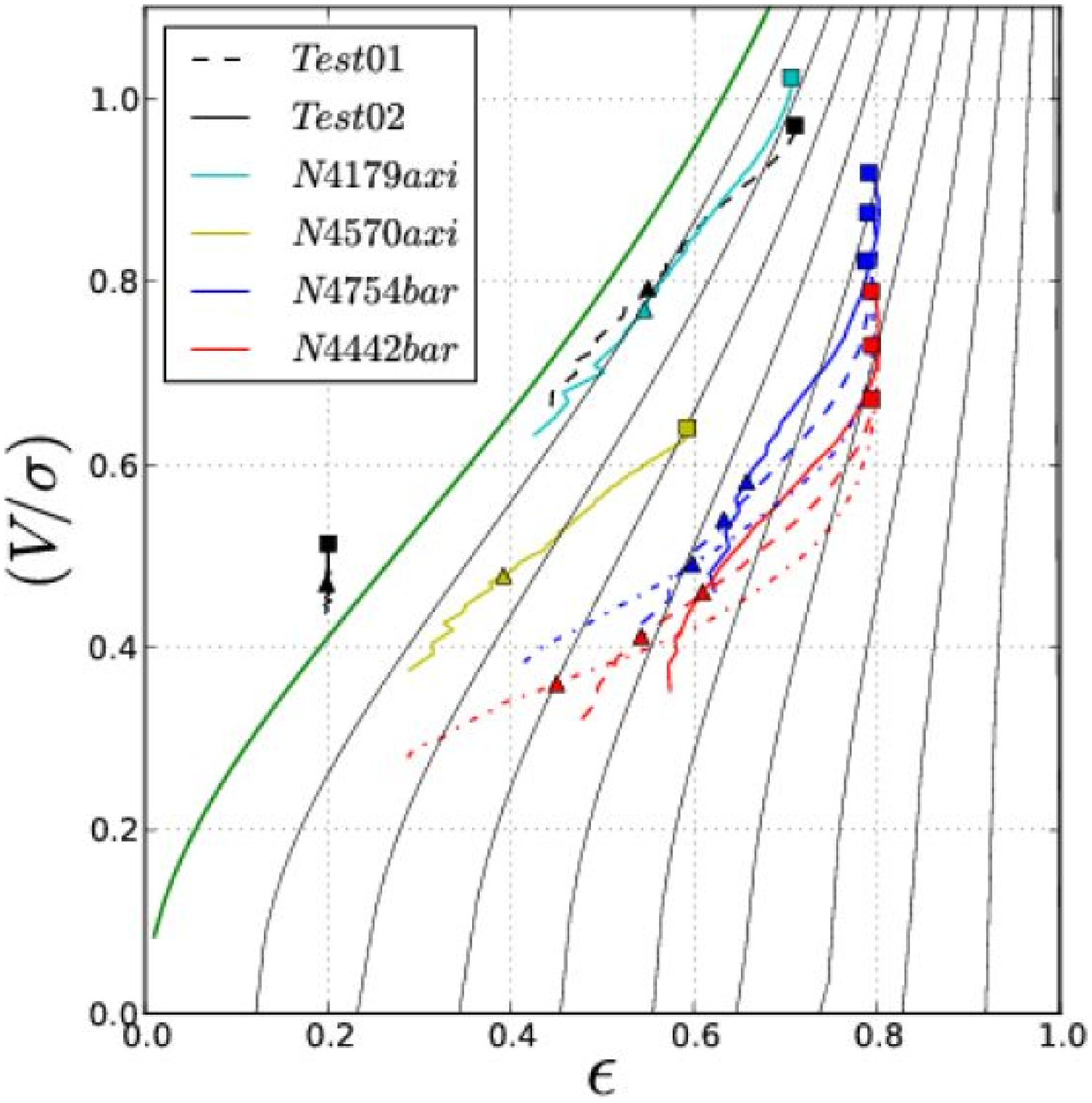}
 \caption{Positions of edge-on projections of all models in the ($V / \sigma$, $\epsilon$) diagram. The two tests models described in Section~\ref{sec:Vsig} are in black. The green line and the thin black solid lines correspond to isotropy and constant anisotropy in the diagram with a step of 0.1 from left to right. The cyan line correponds to $N4179axi$; the yellow one to $N4570axi$ ; the blue ones to $N4754bar$ and the red ones to $N4442bar$. For barred simulations the solid line refers to PA$_{\rm bar}=18^{\circ}$; the dashed one to PA$_{\rm bar}=45^{\circ}$ and the dash-doted one to PA$_{\rm bar}=87^{\circ}$ (we removed the case PA$_{\rm bar}=60^{\circ}$ for better legibility). The wider FOV is represented by a square and our reference FOV in the present study by a triangle. \label{fig:vsigma}}
\end{figure}

To further investigate the impact of the ellipticity $\epsilon$ on $\beta_{z}^{\rm JAM}$, we computed the positions of our mock observations in the ($V / \sigma$, $\epsilon$) diagram \citep{Bin05}.
($V / \sigma$) and $\epsilon$ are computed in an ellipse of area $A = \pi R_e^2$ where $R_e$ is the radius of a cylinder enclosing half of the galaxy light.
In Figure~\ref{fig:vsigma} we present the result for near edge-on projections with $R$ going from $0.5 \times R_e$ to $5.0 \times R_e$.
We only consider the edge-on case in this figure as all our models follow the inclination law \citep[see Sec.~4.3 of][]{BT87}.
For comparison additional simple models were also constructed and projected edge-on.
The first one (named Test01) is based on a MGE parametrization with constant $\epsilon = 1 - q$ and isotropic kinematics (black solid line on the Figure~\ref{fig:vsigma}).
The second one (Test02) is also isotropic but with a non constant ellipticity in the mass distribution, $\epsilon$ increasing with radii (black dashed line). \\

These two test models highlight the fact that when the considered area A is increased ($V / \sigma$) increases as to roughly follow the constant anisotropy lines.
The axisymmetric model $N4179axi$ also lay on a constant anisotropy line although its dynamical structure presents a $\beta_z$ gradient.
Barred simulations present a different behavior depending on PA$_{\rm bar}$.
With broad field of view (FOV) the only effect of the bar is seen through ($V / \sigma$) which decreases when PA$_{\rm bar}$ increases.
But when the size of the FOV is of the order of the bar size, $\epsilon$ is affected by PA$_{\rm bar}$ and then the projections spread over a wide range of anisotropy.
The intrinsic ellipticity of the MGE parametrization plays here an important role for the anisotropy recovery.

\begin{figure}
 \includegraphics[width=\columnwidth]{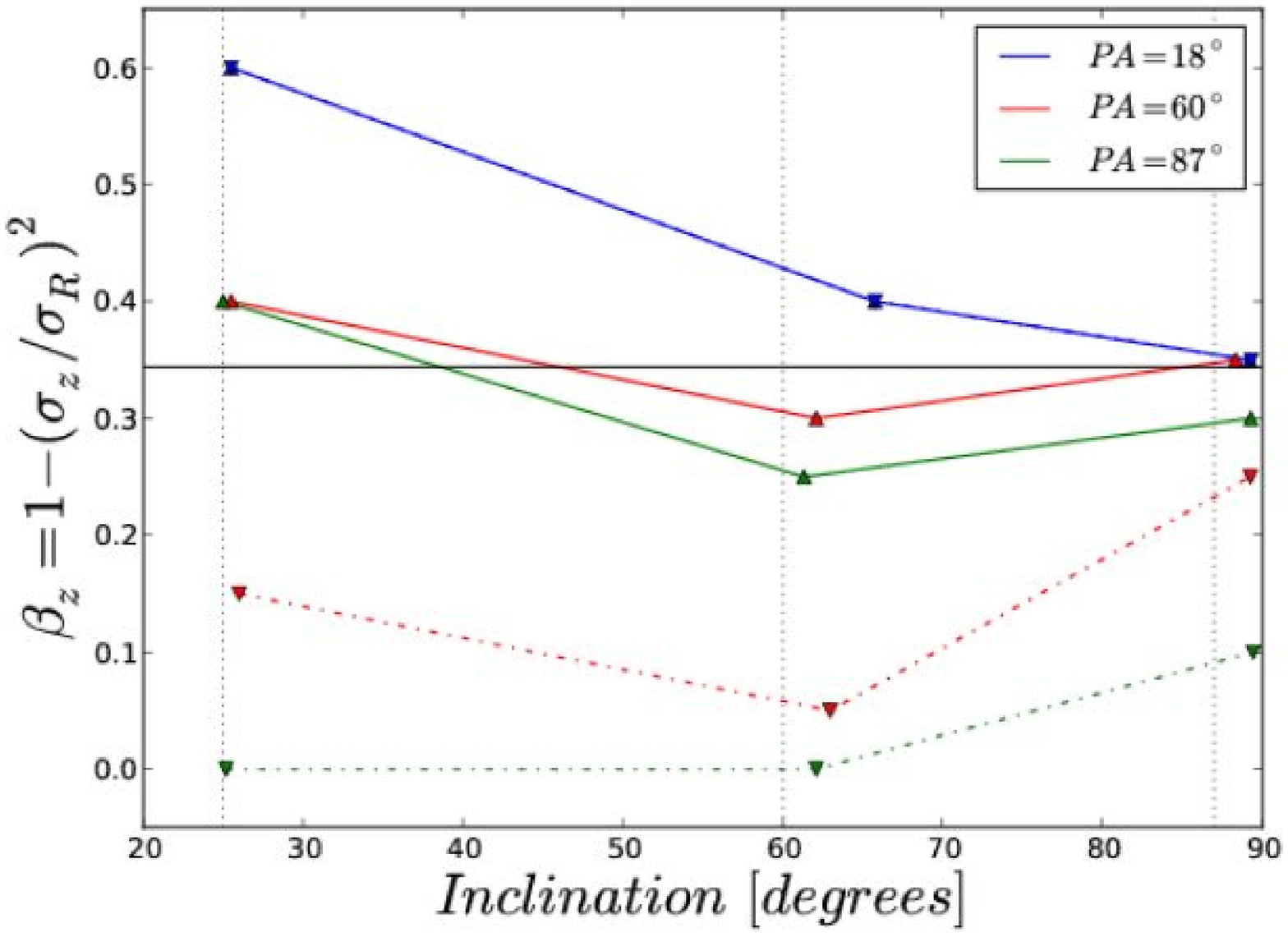}
\caption{Recovered values of the global anisotropy as a function of the projection inclination for two different MGE parametrization of the same model. The color solid lines are the values recovered with a model with forced axis ratio for the MGE parametrization and the dash-dot lines are for a free MGE parametrization (see Section~\ref{sec:Vsig}). Note that the blue solid line and the blue dash-dot line are overlaping. The horizontal black line is the value $\beta_{z}^{\rm SIM}$ computed from the model. \label{fig:QBOUNDS}}
\end{figure}

We made a second test to better understand the effect of the MGE parametrization on a model of barred object.
From the same projection we created two different MGE models : the first one is a "free" model with gaussian axis ratios left unconstrained ; for the second one we forced the maximum and the minimum axis ratio as we do know the intrinsic mass distribution.
We did not include here the case PA$_{\rm bar}=45^{\circ}$ for which the MGE parametrization is forced to be neither flatter nor rounder than the axisymmetric case.
As expected, when we forced the axis ratio during the MGE parametrization, the global anisotropy $\beta_{z}^{\rm JAM}$ recovered was found to be much closer to $\beta_{z}^{\rm SIM}$ as shown in Fig~\ref{fig:QBOUNDS}.
We can then assume that the accuracy in the recovery of the global anisotropy is mainly biased by the MGE parametrization of the photometry.\\

Unfortunately, we cannot always objectively choose the best deprojected model when we apply the JAM modeling method to real observations.
And for barred galaxies the accuracy on $\beta_z$ will only be improved if we can really see the bar or have strong evidence for its presence.

\subsection{Recovery of $M/L$}\label{sec:ML}

\begin{figure}
 \includegraphics[width=\columnwidth]{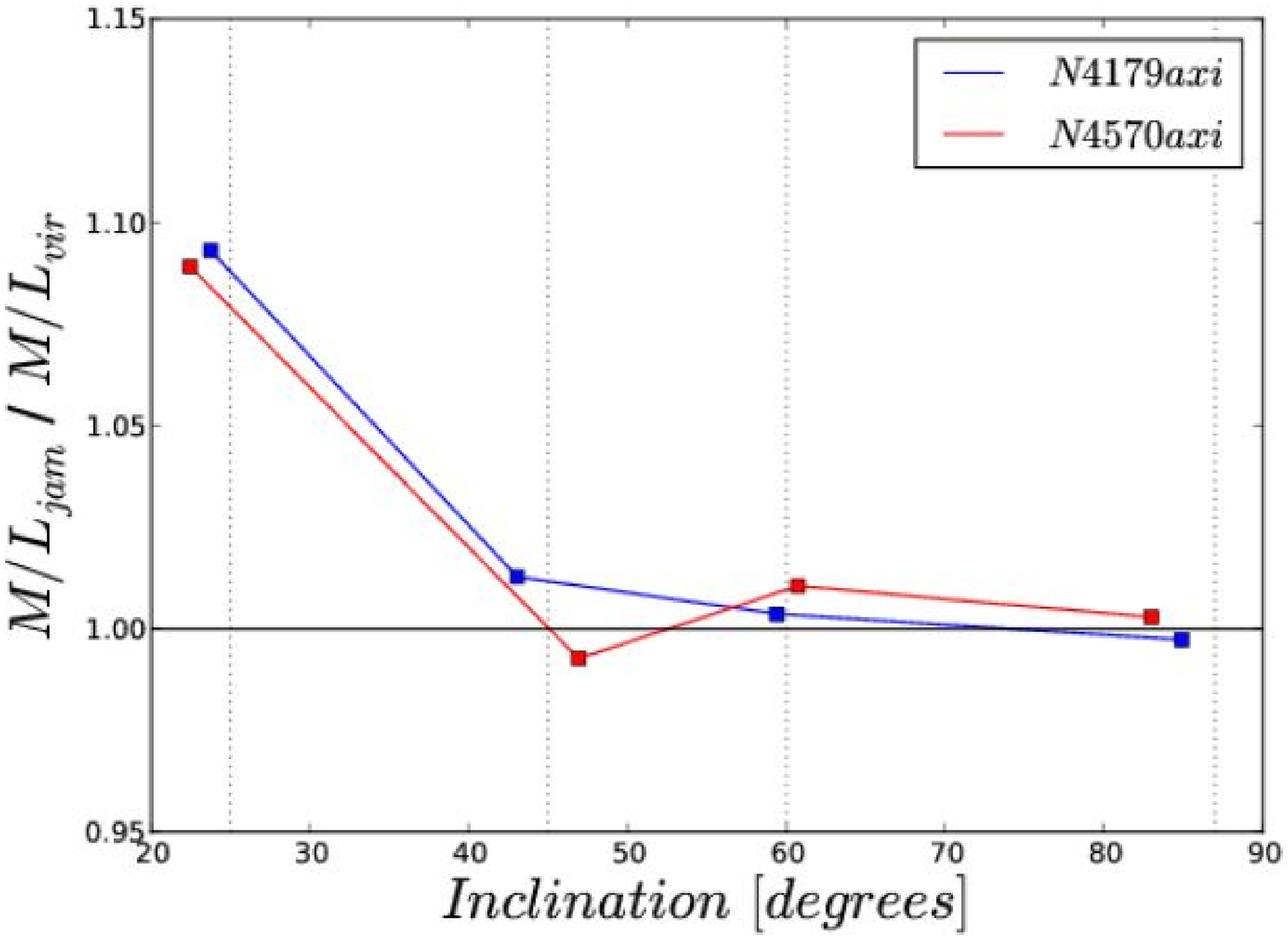}
 \caption{Recovered value of $M/L$ for non barred simulations $N4179axi$ and $N4570axi$ as a function of the inclination of projection. In these regular axisymmetric simulations we find an error of 1.5\% on $M/L$ (excluding the near face-on projection). \label{fig:AxiML}}
\end{figure}

The Mass-to-Light ratio $M/L$ is the third parameter (after the inclination $i$ and the global anisotropy $\beta_z$) computed with the JAM modeling method.

When using the initial conditions, we simply compare the recovered $M/L$ with the input ones.
For evolved galaxies we used instead, as our reference, the $M/L$ computed from the direct application of the virial relation $2K+W=0$ to the simulation particles, where K is the total kinetic energy and W is the total potential energy of our simulations.The relation makes no other assumption that a steady state, and thus provides the natural benchmark against which to compare stationary dynamical models.
In general one expects simulations and real galaxies to satisfy the relation quite accurately, so that the virial $M/L_{\rm vir}$ will agree with the input one $M/L_{\rm SIM}$ and no distinction needs to be made.
However \cite{Thom07} found that  $M/L_{\rm vir}$ can differ from the input one at the 5\% level, due to non stationarity, and our results agree with theirs.
As we are not interested on investigating the stationarity of the model, but only the biases of the modelling method, for maximum accuracy we use as reference $M/L_{\rm vir}$ in all the comparisons which follow.

To probe the robustness of our method we applied it to the four Hernquist particle realisations created from the analytic formula of \cite{Hern90}.
The results both in the recovery of the global anisotropy and the $M/L$ were excellent, with an accurate recovery of $\beta_z$ and errors on $M/L$ of less than 1.5\%. Whilst these simulations are basic and do not reproduce the complexity of a real galaxy, this is a reassuring test of our machinery.

An intermediate case between simple analytic simulations and real galaxies, is the $N4754ini$ model which is just a regular axisymmetric rotating galaxy with a constant anisotropy.
Its mass distribution corresponds to a real galaxy but its intrinsic dynamics are simple, as the velocity anisotropy is set to be constant throughout the galaxy.
The global anisotropy is well recovered within 0.025 and the mass-to-light ratio is recovered with an error of less than 1.5\%. Although unrealistic, this case helps isolate the influence of a variable anisotropy on the results of the JAM modeling method.

\begin{figure}
 \includegraphics[width=\columnwidth]{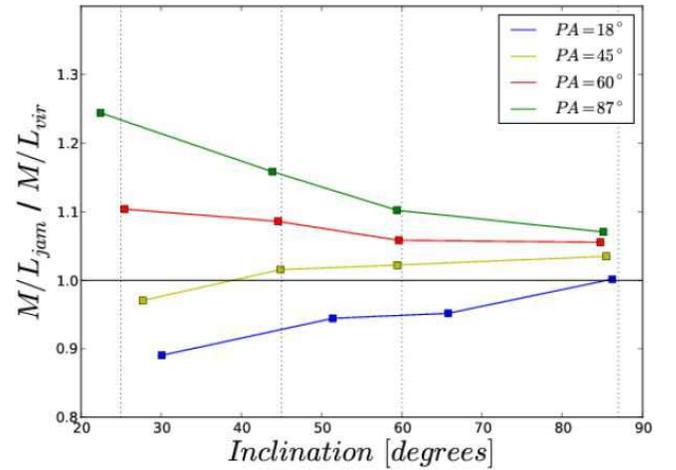}
 \caption{Recovered value of $M/L$ for $N4754bar$ as a function of the projection inclination for different values of PA$_{\rm bar}$. \label{fig:4754bML}}
\end{figure}

\begin{figure}
 \includegraphics[width=\columnwidth]{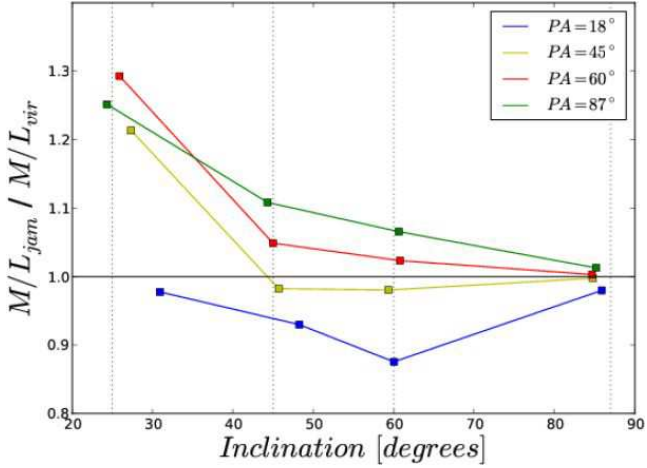}
 \caption{Recovered value of $M/L$ for $N4442bar$ as a function of the projection inclination for different values of PA$_{\rm bar}$. \label{fig:4442bML}}
\end{figure}

We then used the JAM modeling method on $N4179axi$ and $N4570axi$ to explore any systematic bias that may be present {\it without} the presence of a bar.
Results are shown in Figure~\ref{fig:AxiML} and JAM fitting in Fig.~\ref{fig:AxiJAM_4179} and Fig.~\ref{fig:AxiJAM_4570}.
Figure~\ref{fig:AxiML} emphasizes that one can expect significant overestimations of $M/L$ up to $\sim 10$\%, when the galaxy is close to face-on.
This important fact is illustrated with an analytic test in Fig.~4 of \cite{Cap06} and with the galaxy NGC0524 in Fig.~A1 there.
Here we confirm it with the present simulations.
In what follows we will focus on the higher inclinations.
However, one should keep in mind that the $M/L$ of nearly face-on galaxies has to be treated with caution.
For $i=45^{\circ}$ ; $i=60^{\circ}$ and $i=87^{\circ}$ then, we find that in regular axisymmetric cases $M/L$ is recovered with a negligible median bias, and a maximum error of just 1.5\%.

Figure~\ref{fig:4754bML} and Figure~\ref{fig:4442bML} respectively present the results of JAM modeling of $N4754bar$ and $N4442bar$.
As for axisymmetric cases, we only consider here inclination projections with $i\geq45^{\circ}$.
For high inclination projections the $M/L$ is recovered within 3\% for a PA$_{\rm bar} = 45^{\circ}$, which represents the average for random orientations.
However the bias in the recovery can reach up to 15\% in our tests cases.
But the main point is that the recovered $M/L$ is correlated with the position angle of the bar, PA$_{\rm bar}$.
The more the bar is seen end-on, the larger the overestimation, due to the larger velocities of the stars moving along the bar, and towards the line-of-sight, with respect to the other directions in the disk plane.
The reason is that the presence of a bar produces a peak in the $V_{\rm rms}$ maps which is not present in the purely axisymmetric case (see Figure~\ref{fig:Panel}).
In order to fit this peak, the JAM model tend to larger $M/L$ values.
Moreover, the amplitude of the peak increases with PA$_{\rm bar}$, which explains why the value of $M/L_{\rm JAM}$ also increases with the position angle of the bar.
The exclusion of the central regions helps to reduce the bias introduced by the bar in $M/L_{\rm JAM}$, but in cases where the bar dominates the whole field of view, we cannot expect to get rid of its influence.

The case when the bar is seen side-on (PA$_{\rm bar}=18^{\circ}$) for $i=45^{\circ}$ ; $i=60^{\circ}$ and $i=87^{\circ}$ is a special configuration for the mass deprojection.
As the flat bar is deprojected as a flattened disk (see Fig~\ref{fig:MGE_deproj_bis}), following the axisymmetric assumption, the JAM model is unable to reproduce the global shape of the $V_{\rm rms}$ map (see Fig~\ref{fig:JAM_4442_45}-~\ref{fig:JAM_4442_87}, Fig~\ref{fig:JAM_4754_45}-~\ref{fig:JAM_4754_87} top left panel).
The error on the recovered mass-to-light ratio can be up to $\sim 10\%$.
In fact the global shape of the $V_{\rm rms}$ map is dictated by the global anisotropy.
The mass-to-light ratio essentially adjusts the fit to the global level of the second velocity moment.

The position of the bar PA$_{\rm bar}=45^{\circ}$ is a useful reference, as it represents the average value for random orientations.
As previously mentioned in this case the MGE parametrization is hardly affected by PA$_{\rm bar}$.
In this configuration the error on $M/L$ does not exceed 3\%, although the $V_{\rm rms}$ map is not reproduced.
Basically for projections with PA$_{\rm bar} < 45^{\circ}$ $M/L$ is expected to be underestimated while for projections with PA$_{\rm bar} > 45^{\circ}$ it is overestimated.

\begin{figure}
 \includegraphics[width=\columnwidth]{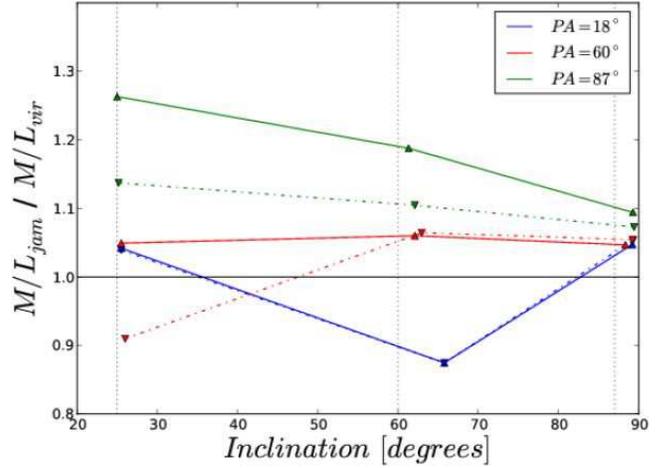}
\caption{Recovered value of the mass-to-light ratio as a function of the projection inclination. The color solid lines are the values recovered with a model with forced axis ratio for the MGE parametrization and the dash-dot lines are for a free MGE parametrization. \label{fig:MLQBOUNDS}}
\end{figure}

Then for PA$_{\rm bar}=60^{\circ}$ and PA$_{\rm bar}=87^{\circ}$, the bar produces a vertically elongated structure in projection and an artificially round bulge when deprojected as an axisymmetric system.
However the reproduction of the $V_{\rm rms}$ shape is still a hard task for the JAM model (see Fig~\ref{fig:JAM_4442_45}-~\ref{fig:JAM_4442_87}, Fig~\ref{fig:JAM_4754_45}-~\ref{fig:JAM_4754_87} right column).
A brief investigation pertaining to the influence of the MGE parametrization on $M/L_{\rm JAM}$, illustrated by Figure~\ref{fig:MLQBOUNDS}, shows that forcing the flattening of the mass distribution in the JAM model does not really affect $M/L_{\rm JAM}$, except for PA$_{\rm bar}=87^{\circ}$.
For this bar position the accuracy in $M/L_{\rm JAM}$ is increased, but at the same time, as previously mentioned in Section~\ref{sec:beta}, the accuracy on the recovered global anisotropy is decreased.
The prediction of the $V$ and $\sigma$ fields, are however significantly improved, when forcing $q_k$, is  as already shown in figure~3 of \cite{Sco09}.

To sum up, when a bar is present the $M/L$ overestimation or underestimation with the JAM modelling method are clearly due : (i) to the fact that the mass is incorrectly deprojected as too flat or too round, when the bar is edge-on or side-on respectively ; and (ii) to the fact that the projected second moments are lower or higher when the bar is edge-on or side-on respectively.

\begin{figure}
 \includegraphics[width=\columnwidth]{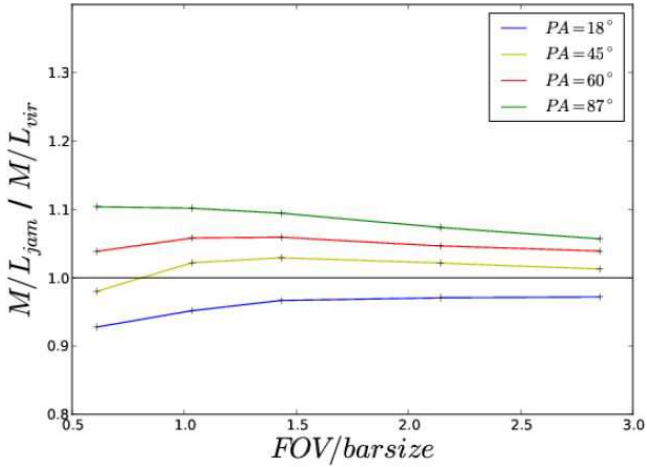}
 \caption{Effect of the FOV on the accuracy of the recovered $M/L$ of $N4754bar$ projected with $i=60^{\circ}$. The $x-axis$ correspond to the size of the side of the field of view normalized by the size of the major axis of the bar determined in Section~\ref{sec:bar}. \label{fig:FOV}}
\end{figure}

To reduce the effect of the bar, one possible solution is to fit the JAM model only to regions where the bar has little or no influence.
We therefore investigated the effect of the size of the field of view (FOV) on $M/L_{\rm JAM}$.
One could expect that when increasing the FOV further we could minimize the effect of the bar on the fit and thus reach a better accuracy in the recovering operation.
To test this, we increased the FOV of our mock data and repeated the JAM fitting procedure (the MGE parametrization was not affected as it is anyway done on a projection of the simulations with a very large field of view).
Figure~\ref{fig:FOV} illustrates $M/L_{\rm JAM}$ as a function of the size of one side of the FOV normalized by the size of the bar for $N4754bar$.
We find that, as expected, the FOV plays a role in the recovered values.
When the size of the FOV exceeds the bar size $\Delta M/L$ decreases and seems to tend to a limit value. This is expected as we cannot totally get rid of the effect of the bar in the $V_{\rm rms}$ maps.
In our study, the typical fitted field of view is quite comparable to the size of the bar itself, meaning that our previous results are close to the worst case scenario.
The relative size of the bar with respect to the size of the field of view is an important ingredient for the recovery of the mass-to-light ratio.

To conclude, when modeling a barred galaxy assuming an axisymmetric mass distribution, the $M/L$ is on average (at PA$_{\rm bar} = 45^{\circ}$) still well recovered.
It can, however, be overestimated, when the bar is parallel to the line-of-sight, or overestimated, when the bar is orthogonal to the line-of-sight, by up to 15\% (in our tests).
The amplitude of these errors mainly depend on the position angle of the bar PA$_{\rm bar}$, but also on the size of the FOV.
Excluding the most central parts and increasing the FOV would naturally tend to reduce the influence of the bar without removing the error on $M/L$ entirely.

\section{Conclusion}\label{sec:Sum}

This paper focuses on the study of the possible biases in the $M/L$ and anisotropy determination for barred simulations, when using axisymmetric dynamical models.
This extends previous studies \citep{Thom07} of triaxial and prolate merger remnants to axisymmetric and barred simulated disk galaxies that better resemble observed fast early-type rotators which constitute the large majority of the gas-poor population in the nearby Universe (see Paper~II and Paper~III).
We do this by generating N-body simulations of objects with properties similar to observed galaxies, both with and without bars.
These are projected at various viewing angles and used to generate mock observations that closely resemble real data.
These data are then fed into the JAM modeling machinery as for real data, the difference being that the intrinsic values of the free JAM model parameters ($i$, $\beta_z$, $M/L$) are known for the simulated data set.

The errors in the recovered inclination increase with inclination due to the fact that, as previously noticed, the models predictions are sensitive to the intrinsic axial ratio of the MGE models.
This implies that for nearly face-on inclinations, where the intrinsic axial ratio of the models change rapidly, the inclination is formally constrained to a fraction of a degree.
This formal accuracy is, however, compensated by a broader degeneracy in the mass deprojection, leading to a small {\em negative} bias in the inclination.
In practice, for the four simulations we constructed (40 different projections in total), the errors never exceeded 5$^{\circ}$.

We confirm previous results that the $M/L$ can be recovered within a few percent when the simulated galaxies are nearly axisymmetric, except for nearly face-on view ($i \lesssim 30^{\circ}$), for which the $M/L$ can be significantly overestimated.
The global anisotropy $\beta_z$ can be difficult to recover, especially at low inclination (near face-on) due to the inclination-anisotropy degeneracy.
This degeneracy implies a significant uncertainty on $\beta_z$ at low inclination, but a smaller error at high inclination.
The global anisotropy is primarily influenced by the flattening (or the roundness) of the MGE parametrization of the projected luminosity.
In the case of regular axisymmetric objects, the main issue is the intrinsic mathematical degeneracy of the luminosity deprojection at low inclination, which affects any axisymmetric deprojection method, including the adopted MGE one.
This results in small deviations of the recovered global anisotropy from the value computed from the numerical simulations: $\beta_z$ is well recovered.

When a bar is present, the mass deprojection becomes the main uncertainty in the models.
The deprojected axisymmetric model will be naturally different from the true non-axisymmetric barred distribution and will change as a function of the observed PA.
Consequently the predicted $V_{\rm rms}$ of the models, as well as the corresponding best fitting $\beta_z$ will change as a function of the PA and can be quite different from the true axially-averaged value.

The mass-to-light ratio is less sensitive to the MGE parametrization than the global anisotropy, but it is biased due to the intrinsic dynamics of the system we want to model.
We find that $M/L$ is mainly influenced by the position of the bar and the size of the field of view.
The error depends upon the position angle of the bar PA$_{\rm bar}$ and can be up to 15\%.
Including only regions far from the bar allows a reduction of the error, but cannot generally avoid it completely.

Our study provides an estimate of the $M/L$ error that can affect the determination of dynamical $M/L$ via axisymmetric models, and in particular using the JAM method.
The large variety of possible shapes, sizes and orientations of bars in galaxies, each with specific dynamics, prevents us from quantifying the exact errors made for individual galaxies.
One should also keep in mind that our study is done on simulations of relatively weak bars.
Therefore, the error on the estimated $M/L$ when modelling galaxies exhibiting stronger bars is expected to be larger.
The objects studied here are still representative \atlas\ fast rotators, this study therefore providing clear guidelines when applying axisymmetric modelling to such large samples.

\section*{acknowledgements}

MC acknowledges support from a Royal Society University Research Fellowship.
This work was supported by the rolling grants `Astrophysics at Oxford' PP/E001114/1 and ST/H002456/1 and visitors grants PPA/V/S/2002/00553, PP/E001564/1 and ST/H504862/1 from the UK Research Councils. RLD acknowledges travel and computer grants from Christ Church, Oxford and support from the Royal Society in the form of a Wolfson Merit Award 502011.K502/jd. RLD also acknowledges the support of the ESO Visitor Programme which funded a 3 month stay in 2010.
SK acknowledges support from the the Royal Society Joint Projects Grant JP0869822.
RMcD is supported by the Gemini Observatory, which is operated by the Association of Universities for Research in Astronomy, Inc., on behalf of the international Gemini partnership of Argentina, Australia, Brazil, Canada, Chile, the United Kingdom, and the United States of America.
TN and MBois acknowledge support from the DFG Cluster of Excellence `Origin and Structure of the Universe'.
MS acknowledges support from a STFC Advanced Fellowship ST/F009186/1.
NS and TAD acknowledge support from an STFC studentship.
(TAD) The research leading to these results has received funding from the European
Community's Seventh Framework Programme (/FP7/2007-2013/) under grant agreement
No 229517.
MBois has received, during this research, funding from the European Research Council under the Advanced Grant Program Num 267399-Momentum.
The authors acknowledge financial support from ESO.


\appendix

\section{Simulations projected density}

We present here the projected density maps for the simulations used in the present study at two different scales.

\begin{figure*}
 \includegraphics[height=0.9\textheight]{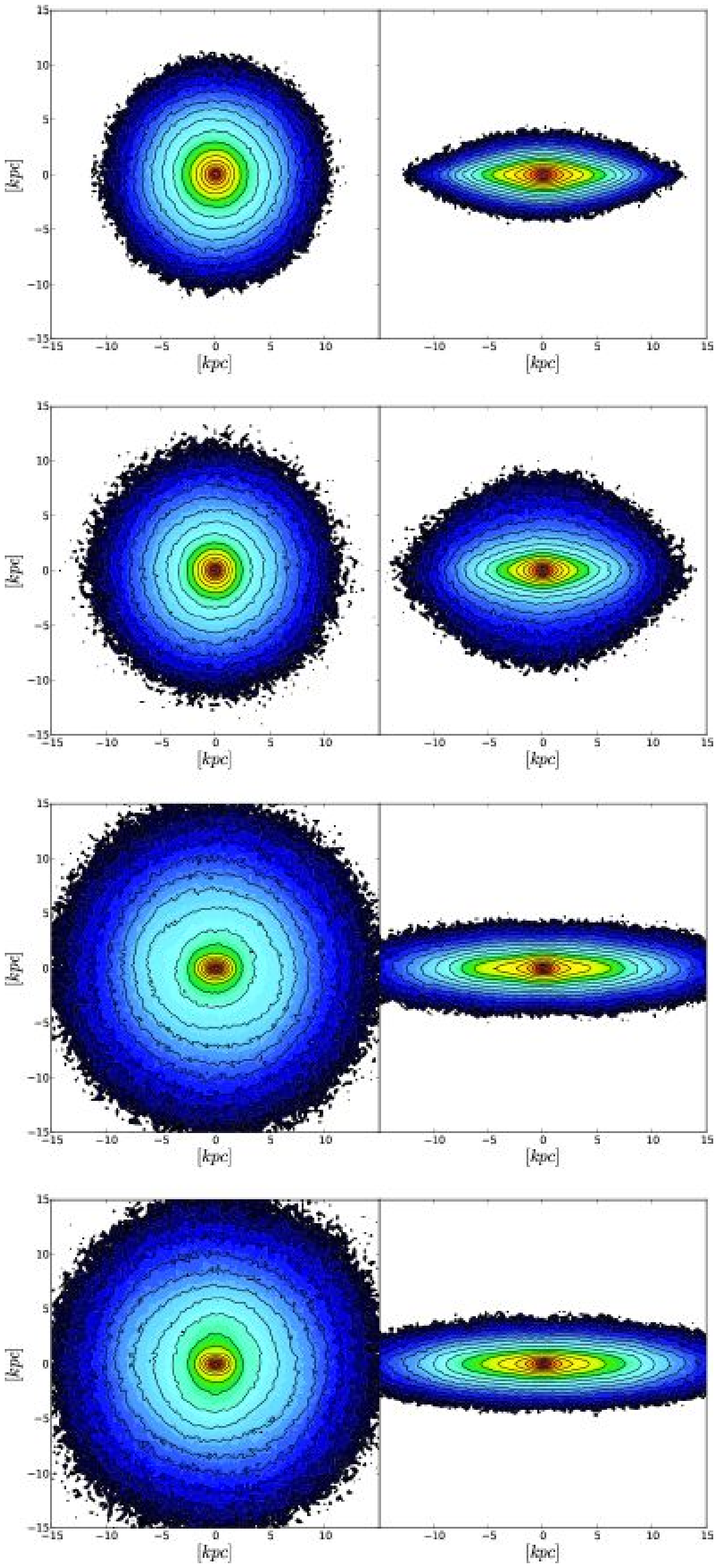}
 \caption{Face-on (left panels) and edge-on (right panels) projections of the final state of the four simulations. From top to bottom : $N4179axi$, $N4570axi$, $N4442bar$ and $N4754bar$.  \label{fig:Simu_final}}
\end{figure*}

\begin{figure*}
 \includegraphics[height=0.9\textheight]{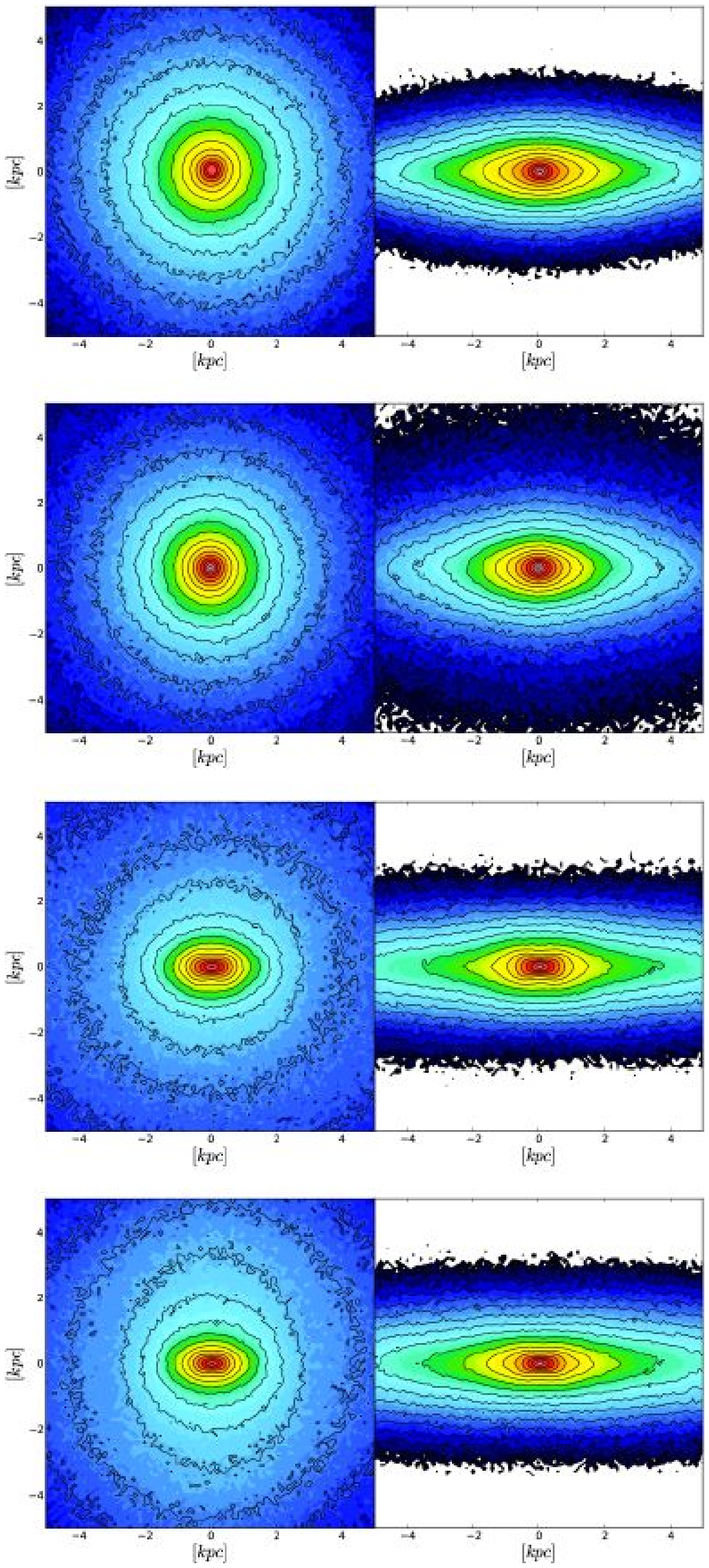}
 \caption{Same as Fig~\ref{fig:Simu_final} with a field of view of $10kpc \times 10kpc$.  \label{fig:Simu_final_2}}
\end{figure*}

\section{MGE deprojection}

We illustrate with the following figure the intrinsic degeneracy present in the derojection process of inclined galaxies.

\begin{figure*}
 \includegraphics[height=0.8\textheight]{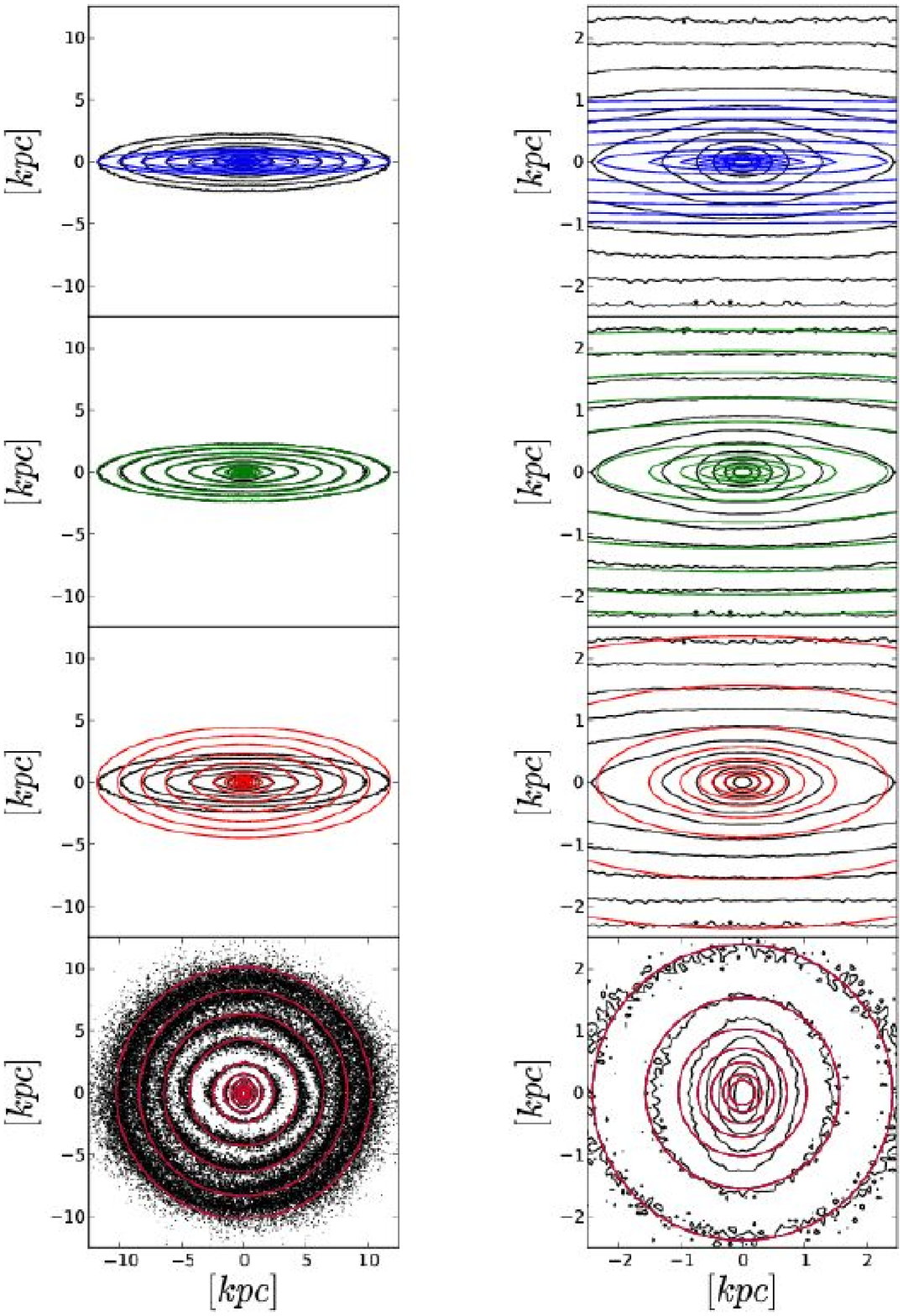}
 \caption{Degeneracy in the edge-on deprojected luminosity with a field of view pf $15kpc \times 15kpc$ (left column) and $5kpc \times 5kpc$ (right column). The first three rows represent the edge-on deprojection of MGE models with an apparent axis ratio $q'=0.907$, $q'=0.910$ and $q'=0.920$ (from top to bottom respectively) when $i=25^{\circ}$, superposed to the edge-on averaged projected luminosity of $N4754bar$ (black contours). In the last row the projected luminosity with $i=25^{\circ}$ is ploted for all the three previous MGE models in addition to the luminosity contours of $N4754bar$ with $i=25^{\circ}$ and $PA_{bar}=87^{\circ}$. We see that in this latter case, even though the projected models are indistinguishable, the deprojected ones can be very different. This is a manifestation of the unavoidable intrinsic degeneracy in the mass deprojection of axisymmetric bodies at low inclination. \label{fig:MGE_degen}}
\end{figure*}

\section{JAM recovery summary}

\begin{table*}
 \begin{tabular*}{0.8\textwidth}{c c c c c c c c c}
  Model & $i_{\rm SIM} $ & PA$_{\rm bar}$ & $i_{\rm JAM}$ & $\beta_{z}^{\rm SIM}$ & $\beta_{z}^{\rm JAM}$ & $M/L_{\rm vir}$ & $M/L_{\rm JAM}$ & $M/L$ Error in \% \\
  \hline
  \hline
  N4179axi & 25 & $\emptyset$ & 23.8 & 0.106 & 0.4 & 6.29 & 6.87 & 9.22 \\
   & 45 & $\emptyset$ & 43.1 & 0.106 & 0.2 & 6.29 & 6.36 & 1.11 \\
   & 60 & $\emptyset$ & 59.4 & 0.106 & 0.05 & 6.29 & 6.30 & 0.16 \\
   & 87 & $\emptyset$ & 84.9 & 0.106 & 0.1 & 6.29 & 6.26 & 0.48 \\
  \hline
  N4570axi & 25 & $\emptyset$ & 22.4 & 0.145 & 0.25 & 16.93 & 18.43 & 8.86 \\
   & 45 & $\emptyset$ & 46.9 & 0.145 & 0.1 & 16.93 & 16.80 & 0.77 \\
   & 60 & $\emptyset$ & 60.7 & 0.145 & 0.1 & 16.93 & 17.10 & 1.00 \\
   & 87 & $\emptyset$ & 82.9 & 0.145 & 0.1 & 16.93 & 16.97 & 0.24 \\
  \hline
  N4442bar & 25 & 18 & 30.9 & 0.344 & 0.25 & 17.99 & 17.58 & 2.28 \\
   & 25 & 45 & 27.3 & 0.344 & 0.55 & 17.99 & 21.82 & 21.29 \\
   & 25 & 60 & 25.9 & 0.344 & 0.6 &  17.99 & 23.25 & 29.24 \\
   & 25 & 87 & 24.3 & 0.344 & 0.35 & 17.99 & 22.51 & 25.13 \\
  \\
   & 45 & 18 & 48.2 & 0.344 & 0.45 & 17.99 & 16.72 & 7.06 \\
   & 45 & 45 & 45.6 & 0.344 & 0.1 &  17.99 & 17.67 & 1.78 \\
   & 45 & 60 & 44.9 & 0.344 & 0.0 &  17.99 & 18.86 & 4.84 \\
   & 45 & 87 & 44.3 & 0.344 & 0.0 &  17.99 & 19.93 & 10.78 \\
  \\
   & 60 & 18 & 60.0 & 0.344 & 0.6 &  17.99 & 15.74 & 12.51 \\
   & 60 & 45 & 59.3 & 0.344 & 0.2 &  17.99 & 17.63 & 2.00 \\
   & 60 & 60 & 60.7 & 0.344 & 0.05 & 17.99 & 18.40 & 2.28 \\
   & 60 & 87 & 60.6 & 0.344 & 0.0 &  17.99 & 19.16 & 6.50 \\
  \\
   & 87 & 18 & 85.8 & 0.344 & 0.25 & 17.99 & 17.62 & 2.06 \\
   & 87 & 45 & 84.7 & 0.344 & 0.25 & 17.99 & 17.94 & 0.28 \\
   & 87 & 60 & 84.6 & 0.344 & 0.2 &  17.99 & 18.04 & 0.28 \\
   & 87 & 87 & 85.1 & 0.344 & 0.1 &  17.99 & 18.21 & 1.22 \\
  \hline
  N4754bar & 25 & 18 & 30.1 & 0.343 & 0.0 & 11.04 & 9.82 & 11.05 \\
   & 25 & 45 & 27.7 & 0.343 & 0.3 & 11.04 & 10.70 & 3.08 \\
   & 25 & 60 & 25.4 & 0.343 & 0.3 & 11.04 & 12.18 & 10.33 \\
   & 25 & 87 & 22.5 & 0.343 & 0.15 & 11.04 & 13.73 & 24.37 \\
  \\
   & 45 & 18 & 51.4 & 0.343 & 0.0 & 11.04 & 10.42 & 5.62 \\
   & 45 & 45 & 44.9 & 0.343 & 0.6 & 11.04 & 11.20 & 1.45 \\
   & 45 & 60 & 44.6 & 0.343 & 0.6 & 11.04 & 11.98 & 8.51 \\
   & 45 & 87 & 43.9 & 0.343 & 0.6 & 11.04 & 12.78 & 15.76 \\
  \\
   & 60 & 18 & 65.8 & 0.343 & 0.0 & 11.04 & 10.50 & 4.89 \\
   & 60 & 45 & 59.4 & 0.343 & 0.6 & 11.04 & 11.28 & 2.17 \\
   & 60 & 60 & 59.6 & 0.343 & 0.05 & 11.04 & 11.68 & 5.80 \\
   & 60 & 87 & 59.4 & 0.343 & 0.0 & 11.04 & 12.16 & 10.14 \\
  \\
   & 87 & 18 & 86.2 & 0.343 & 0.2 & 11.04 & 11.05 & 0.09 \\
   & 87 & 45 & 85.5 & 0.343 & 0.2 & 11.04 & 11.42 & 3.44 \\
   & 87 & 60 & 84.7 & 0.343 & 0.2 & 11.04 & 11.64 & 5.43 \\
   & 87 & 87 & 85.1 & 0.343 & 0.1 & 11.04 & 11.81 & 6.97 \\
  \hline
 \end{tabular*}
 \centering
 \caption{Table summarizing the values of $i$, PA$_{\rm bar}$, $\beta_{z}$ and $M/L$ of our mock observations and the values recovered by the JAM modelling method. \label{tab:sum}}
\end{table*}

\section{JAM fitting}

\begin{figure*}
 \includegraphics[width=\textwidth]{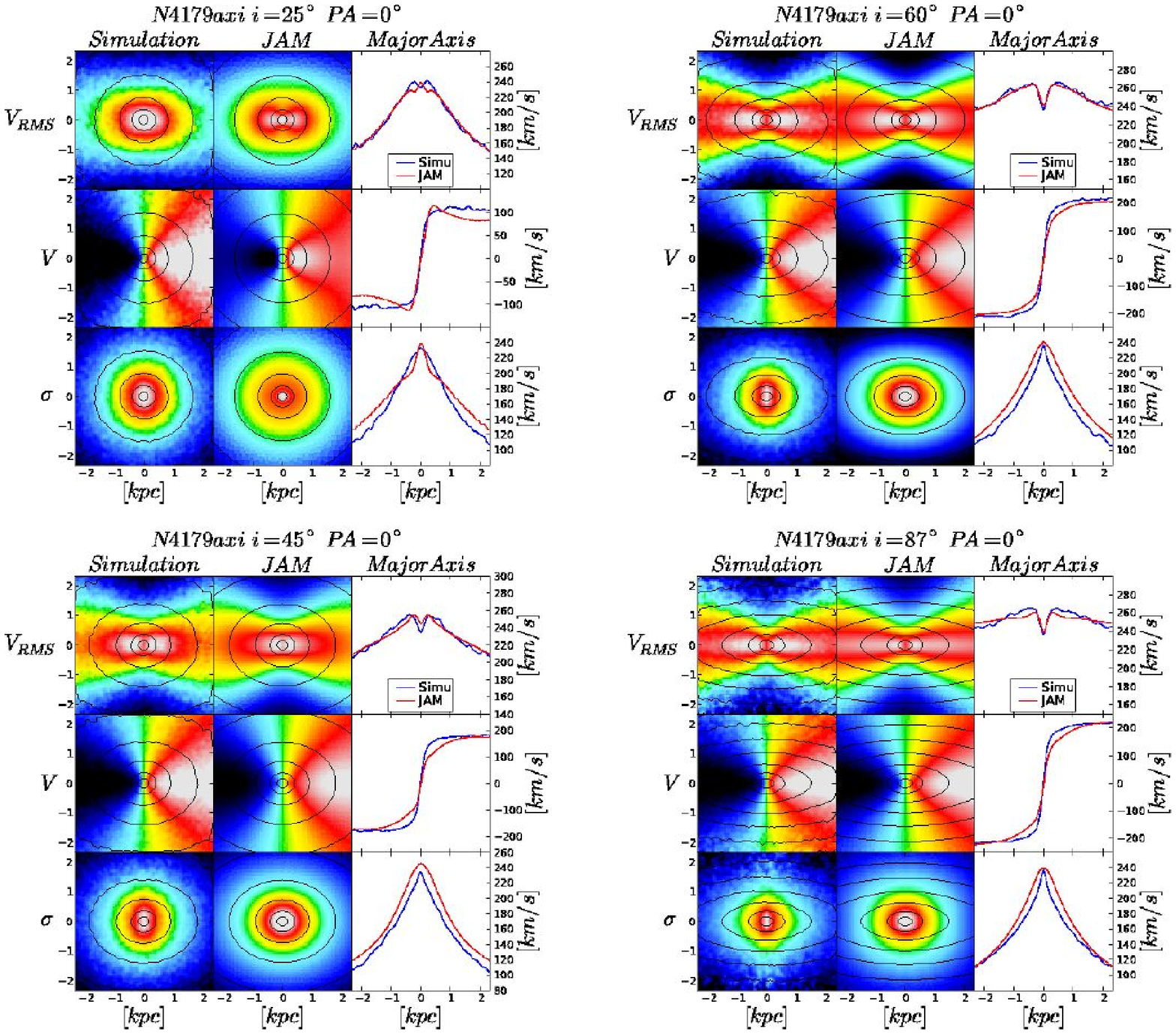}
 \caption{Comparison between the simulations projected velocity maps and the best JAM fitting for $N4179axi$ for the four angles of projection.  \label{fig:AxiJAM_4179}}
\end{figure*}

\begin{figure*}
 \includegraphics[width=\textwidth]{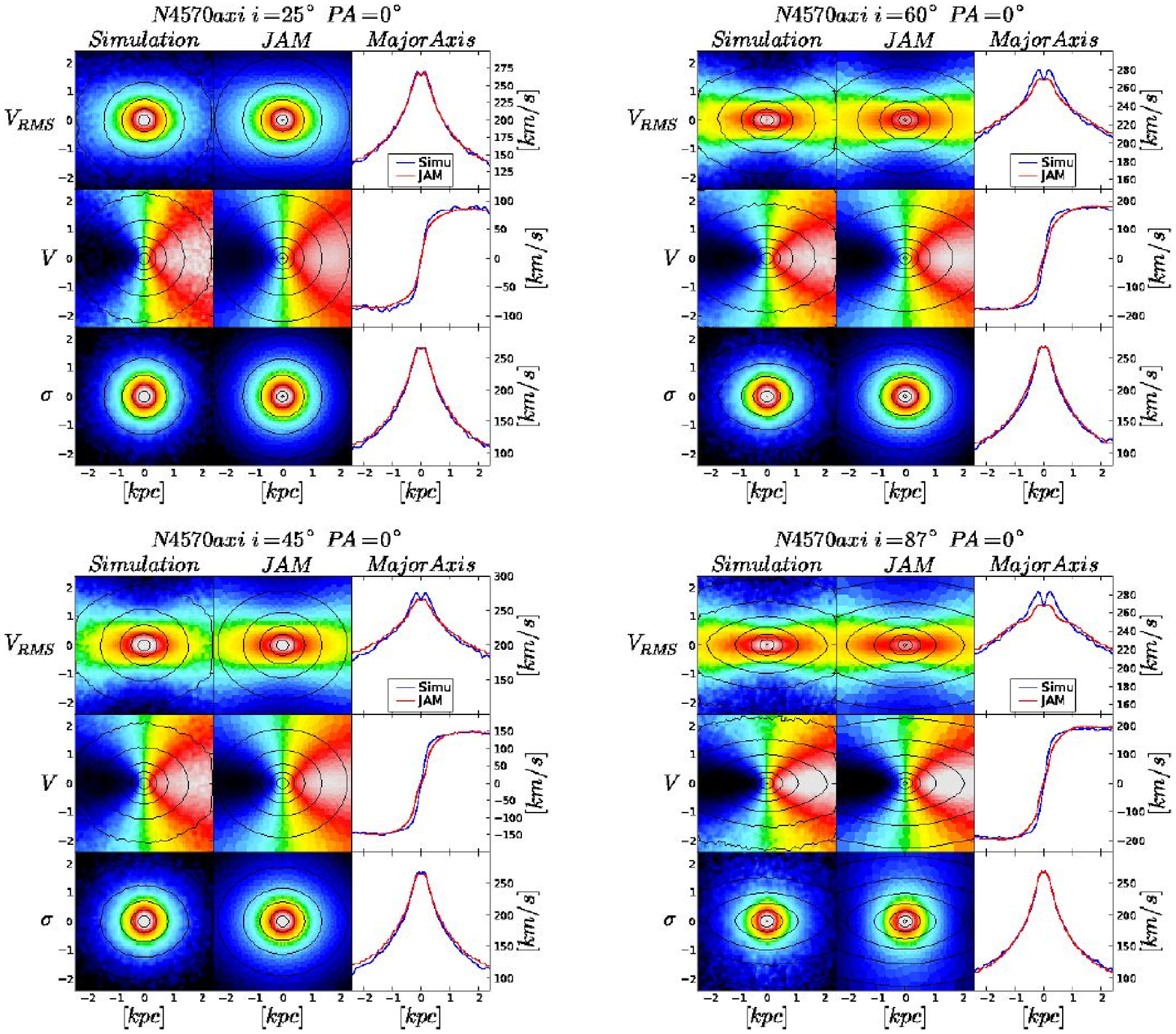}
 \caption{Comparison between the simulations projected velocity maps and the best JAM fitting for $N4570axi$ for the four angles of projection.  \label{fig:AxiJAM_4570}}
\end{figure*}

\begin{figure*}
 \includegraphics[width=\textwidth]{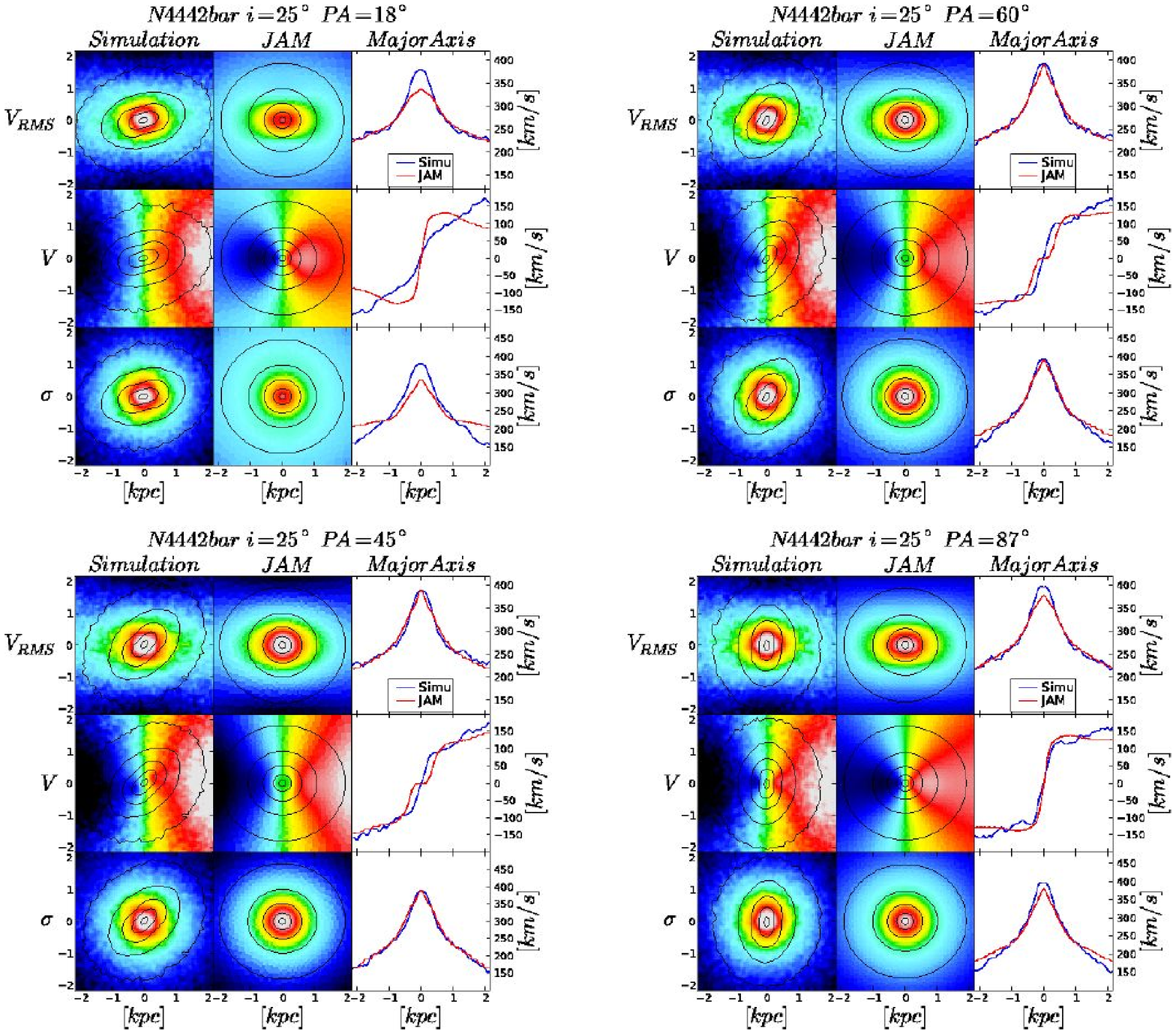}
 \caption{Comparison between the simulations projected velocity maps and the best JAM fitting for $N4442bar$ for $i=25^{\circ}$ and the four PA$_{\rm bar}$.  \label{fig:JAM_4442_25}}
\end{figure*}

\begin{figure*}
 \includegraphics[width=\textwidth]{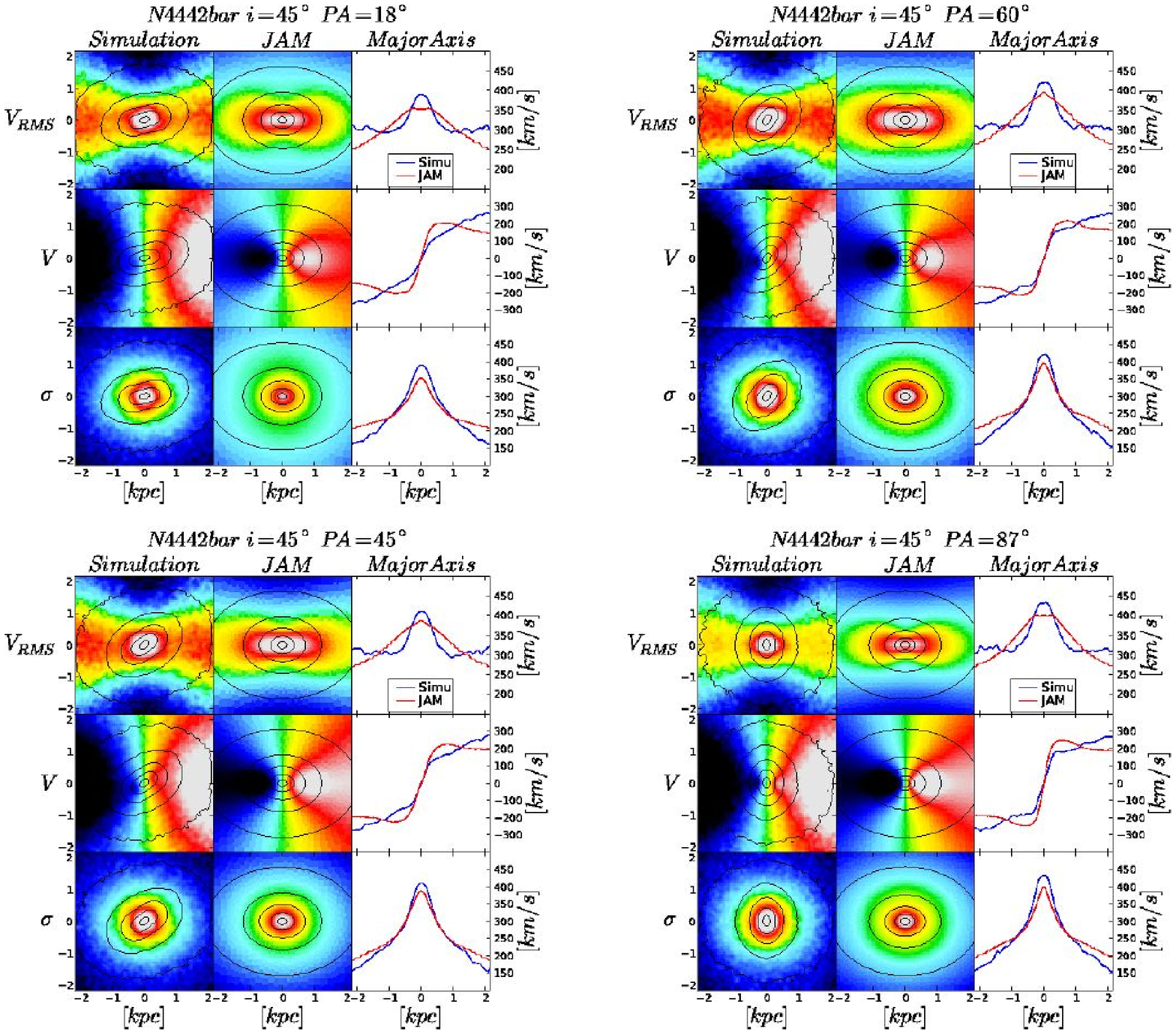}
 \caption{Comparison between the simulations projected velocity maps and the best JAM fitting for $N4442bar$ for $i=45^{\circ}$ and the four PA$_{\rm bar}$.  \label{fig:JAM_4442_45}}
\end{figure*}

\begin{figure*}
 \includegraphics[width=\textwidth]{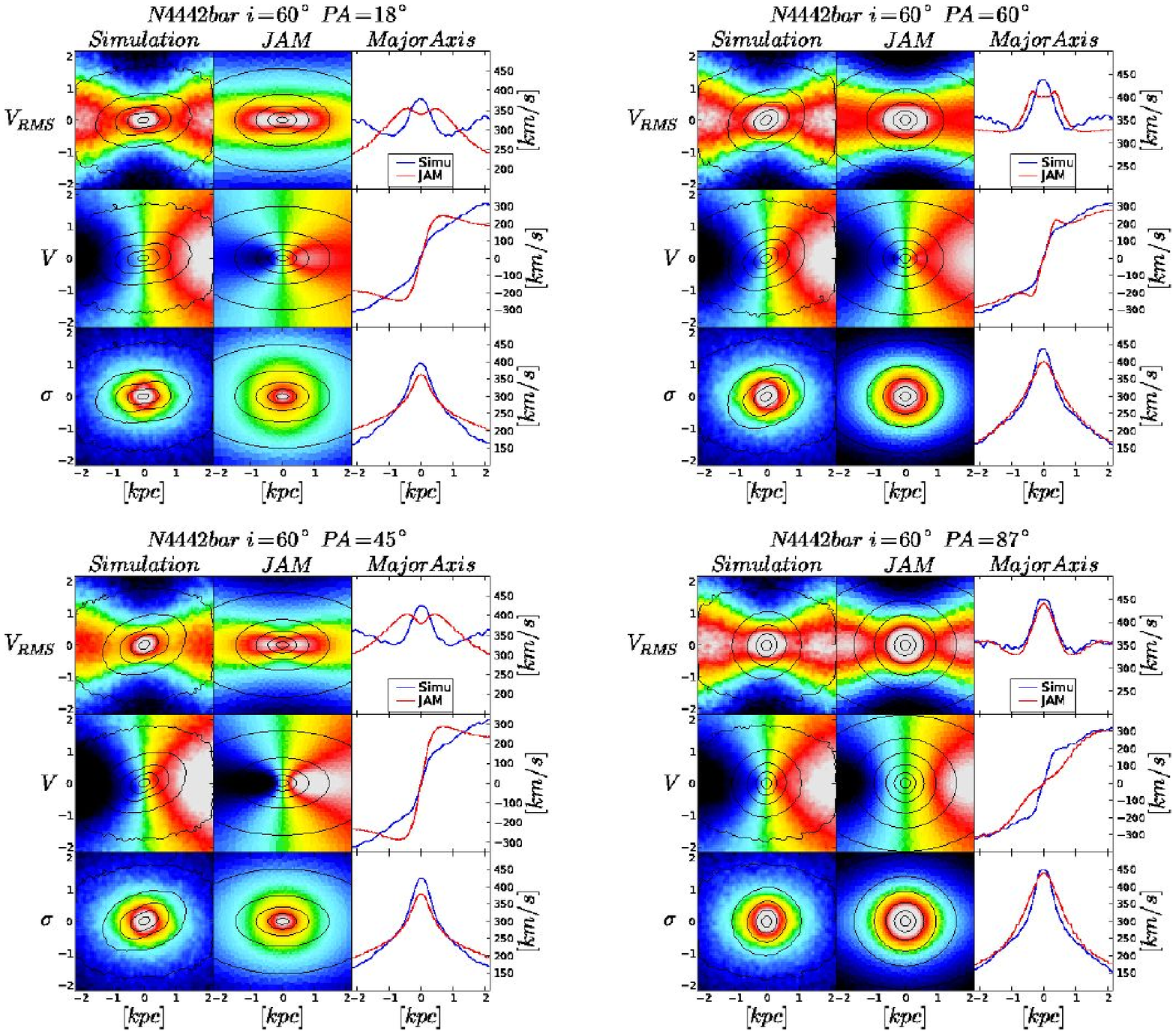}
 \caption{Comparison between the simulations projected velocity maps and the best JAM fitting for $N4442bar$ for $i=60^{\circ}$ and the four PA$_{\rm bar}$.  \label{fig:JAM_4442_60}}
\end{figure*}

\begin{figure*}
 \includegraphics[width=\textwidth]{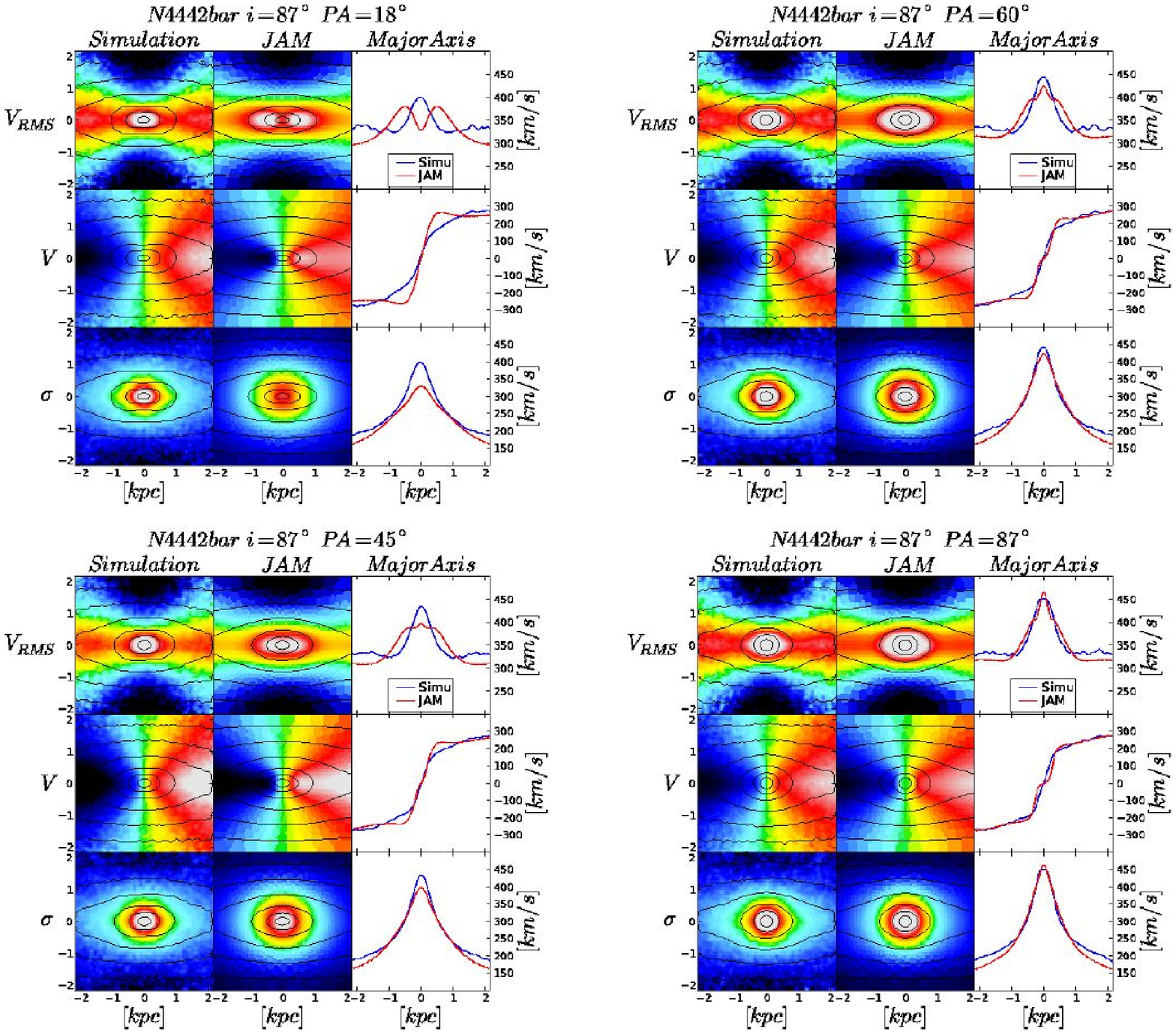}
 \caption{Comparison between the simulations projected velocity maps and the best JAM fitting for $N4442bar$ for $i=87^{\circ}$ and the four PA$_{\rm bar}$.  \label{fig:JAM_4442_87}}
\end{figure*}

\begin{figure*}
 \includegraphics[width=\textwidth]{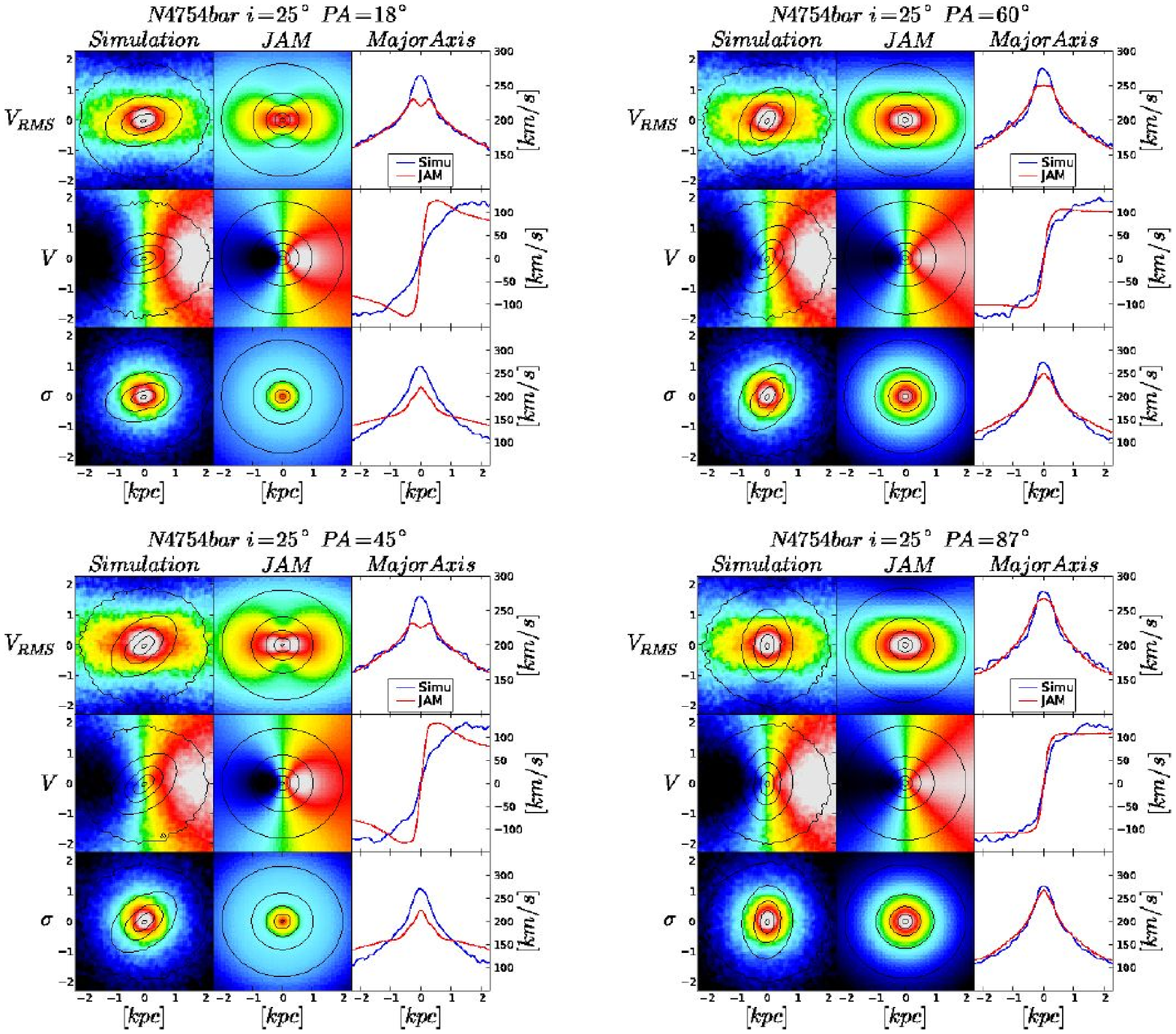}
 \caption{Comparison between the simulations projected velocity maps and the best JAM fitting for $N4754bar$ for $i=25^{\circ}$ and the four PA$_{\rm bar}$.  \label{fig:JAM_4754_25}}
\end{figure*}

\begin{figure*}
 \includegraphics[width=\textwidth]{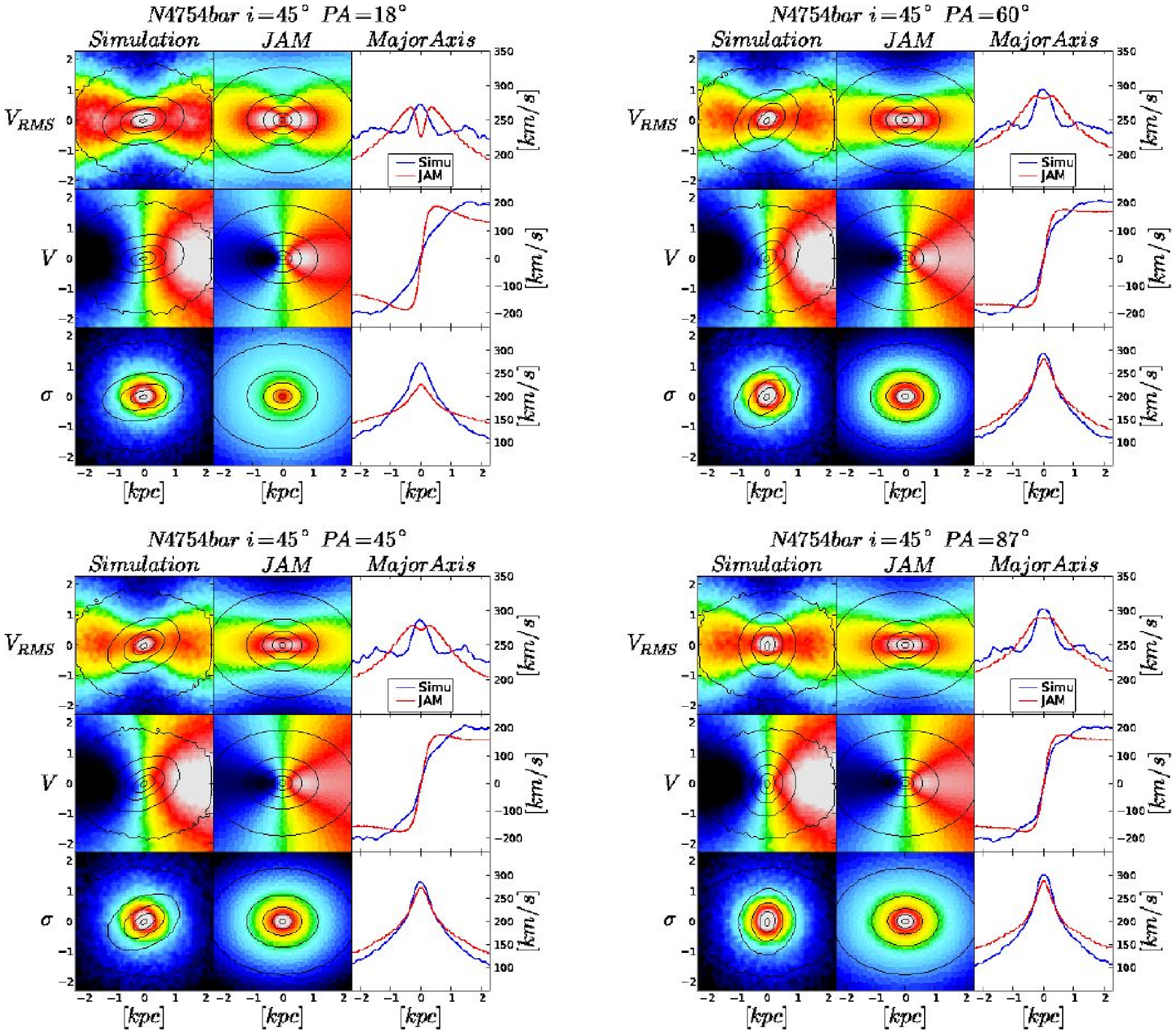}
 \caption{Comparison between the simulations projected velocity maps and the best JAM fitting for $N4754bar$ for $i=45^{\circ}$ and the four PA$_{\rm bar}$.  \label{fig:JAM_4754_45}}
\end{figure*}

\begin{figure*}
 \includegraphics[width=\textwidth]{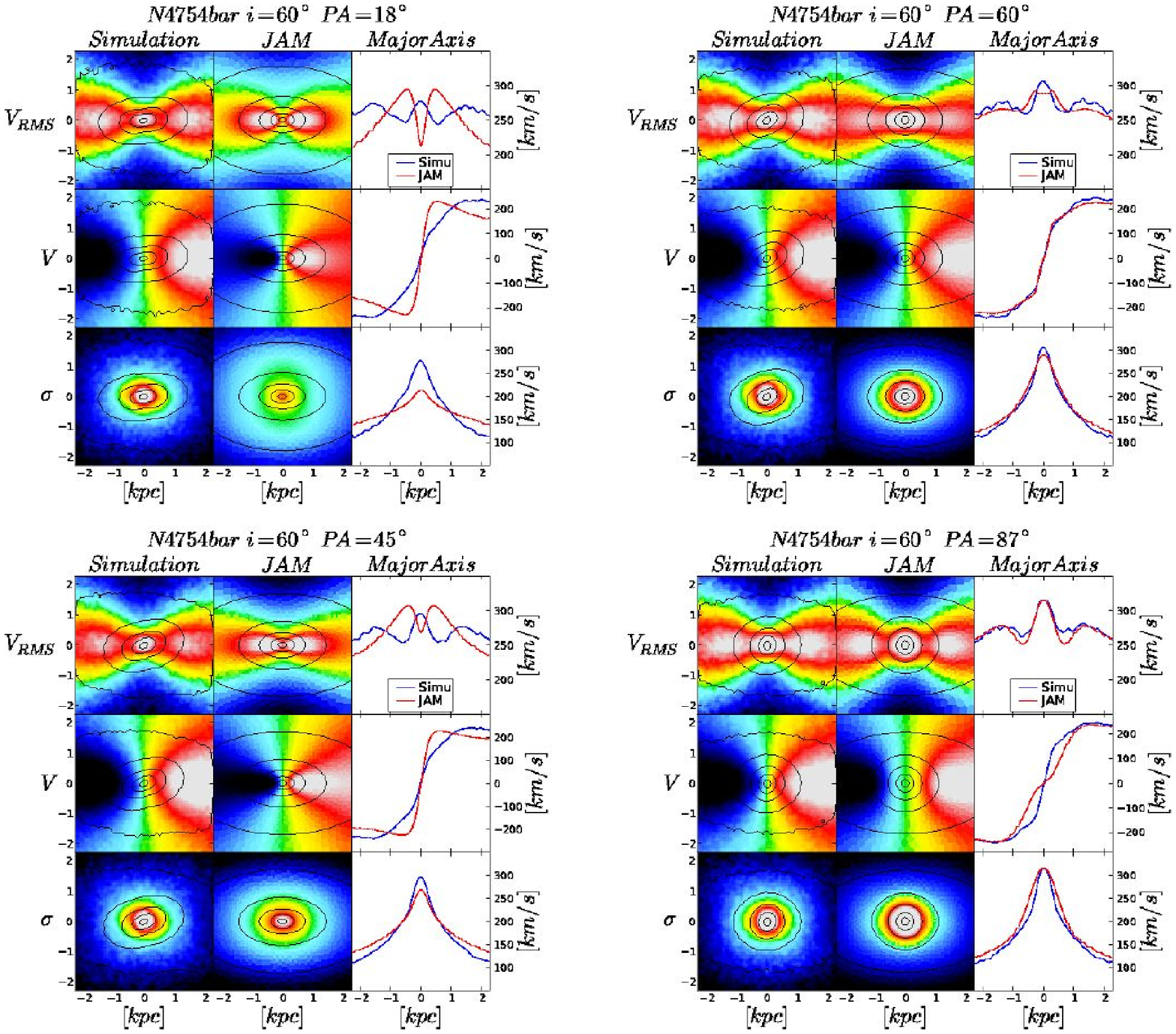}
 \caption{Comparison between the simulations projected velocity maps and the best JAM fitting for $N4754bar$ for $i=60^{\circ}$ and the four PA$_{\rm bar}$.  \label{fig:JAM_4754_60}}
\end{figure*}

\begin{figure*}
 \includegraphics[width=\textwidth]{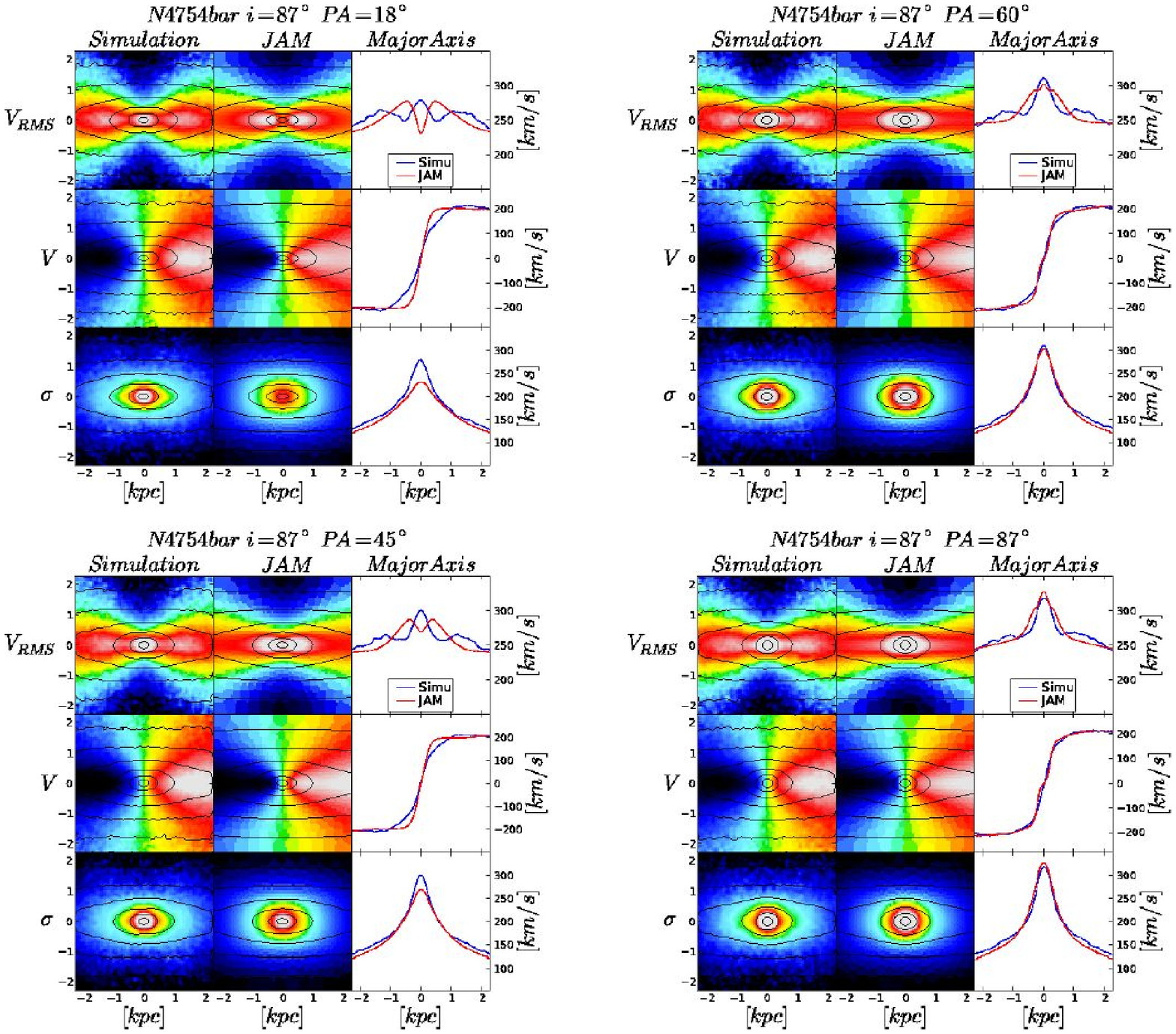}
 \caption{Comparison between the simulations projected velocity maps and the best JAM fitting for $N4754bar$ for $i=87^{\circ}$ and the four PA$_{\rm bar}$.  \label{fig:JAM_4754_87}}
\end{figure*}

\label{lastpage}

\end{document}